%% file: main.tex
\documentclass[journal]{IEEEtran}
\input{preamble}
\usepackage{graphicx}\graphicspath{{./figs/}}
\usepackage{cite}
\ifCLASSOPTIONcompsoc
	\usepackage[caption=false,font=normalsize,labelfont=sf,textfont=sf]{subfig}
\else
	\usepackage[caption=false,font=footnotesize]{subfig}
\fi
\usepackage{color}
\newcommand{\blue}[1]{{\color{black}#1}}

\usepackage[normalem]{ulem}
\usepackage{booktabs}

\begin{document}

\title{Low-Complexity Integer-Forcing Methods for Block Fading MIMO Multiple-Access Channels}

\author{Ricardo Bohaczuk Venturelli and Danilo Silva\thanks{The Ad Hoc Associate Editor coordinating the review of this manuscript and approving it for publication was Prof. Cristiano Magalh\~{a}es Panazio.}\thanks{Ricardo Bohaczuk Venturelli and Danilo Silva are with the Department of Electrical and Eletronic Engineering, Federal University of Santa Catarina, Florianopolis-SC, Brazil (e-mails: ricardo.bventurelli@gmail.com, danilo.silva@ufsc.br).}\thanks{This work was partially supported by 
CAPES under Grant 1345815 and by 
CNPq under Grants 153535/2016-4, 429097/2016-6 and 310343/2016-0.}\thanks{A preliminary version of this paper was presented at the XXXIV Simp\'osio
	Brasileiro de Telecomunica\c{c}\~{o}es e Processamento de Sinais (SBrT'16), Santar\'em, PA, Brazil, August 30-September 4, 2016 \cite{Venturelli.2016:LowComplexityIFforBlockFading}.}
\thanks{Digital Object Identifier: 10.14209/jcis.2017.14} }

\maketitle

\begin{abstract}
Integer forcing is an alternative approach to conventional linear receivers for multiple-antenna systems. 
In an integer-forcing receiver, integer linear combinations of messages are extracted from the received matrix before each individual message is recovered.
Recently, the integer-forcing approach was generalized to a block fading scenario. Among the existing variations of the scheme, the ones with the highest achievable rates have the drawback that no efficient algorithm is known to find the best choice of integer linear combination coefficients. In this paper, we propose several sub-optimal methods to find these coefficients with low complexity, 
covering
both parallel and successive interference cancellation versions of the receiver.
Simulation results show that the proposed methods attain a performance close to optimal in terms of achievable rates for a given outage probability.
Moreover, a low-complexity implementation using root LDPC codes is developed, showing that the benefits of the proposed methods also carry on to practice.

\end{abstract}

\section{Introduction} \label{sec:introduction}

Integer-forcing (IF) receivers are an alternative to conventional methods of equalization, such as zero-forcing (ZF) and minimum-mean-squared-error (MMSE) equalization \cite{Zhan.2014:IFLinearReceivers} for multiple-input and multiple-output (MIMO) channels. 
The IF approach follows from the compute-and-forward framework \cite{Nazer.2011:CaFHarnessing,Feng.2013:AlgebricApproach} for relay networks, where the receivers attempt to extract integer linear combinations of the transmitted messages from the received signals, before recovering the messages themselves. 

Although joint maximum likelihood (ML) receivers achieve the best performance \blue{among} all methods, by searching over all possible transmitted codewords \cite{Tse.2005:FundamentalsWireless}, their complexity is prohibitively high, increasing exponentially with the number of users. In contrast, IF receivers have a much lower complexity and can approach the ML performance in many situations \cite{Zhan.2014:IFLinearReceivers}. Moreover, the performance of an IF receiver can be further improved in some situations by successive computation, leading to the so-called successive IF (SIF) receiver, analogously to the successive interference cancellation (SIC) technique for conventional linear receivers.

\blue{Recent works on IF include the design of practical channel codes compatible with IF \cite{Ordentlich.2010:AchievingGainsbyIFwithBinaryCodes,Chae.2017:MultilevelCodingScheme}, efficient methods to select the coefficients of the integer linear combinations \cite{Ding.2015:ExactSMPAlgorithmIFMIMOReceivers,Liu.2016:EfficientIntegerSearch}, as well as its application to relay networks \cite{Azimi.2016:IFandForwardforMIMOTwoWayRelay}. Moreover, the IF principle has been used not only as receive method in the MIMO uplink scenario, but also for precoding in the MIMO downlink scenario \cite{He.2014:UplinkDownlinkDualityforIF,Silva.2017:IFPrecodingBroadcastChannel} and even for source coding \cite{Ordentlich.2017:IFsourceCoding}}.

The main results about IF \blue{receivers} consider static fading, where all symbols of a codeword are subject to the same channel fading. However, in a practical situation where a powerful code with large blocklength is used, it may not be realistic to assume that all symbols of the codeword are subject to the same channel fading. Therefore, channels that allow block fading \cite{Knopp.2000:CodingforBlockFading}, where the channel fading can vary during the transmission of a codeword, seem to be a more realistic model. 

In a recent work, El Bakoury and Nazer \cite{Bakoury.2015:ImpactofChannelVariation} generalize the IF approach to a block fading scenario. They described two decoding methods for block fading, which are called AM (arithmetic mean) and GM (geometric mean) decoding. The AM decoding method approximates the effective noise seen on all blocks (after equalization) as having the same variance, as in the case of static fading. On the other hand, the GM decoding method optimally exploits the diversity inherent in the channel variation, allowing to achieve higher rates than with AM decoding. Both decoding methods are applicable to both  (non-successive) IF and SIF receivers.

The rates achievable by all these receivers depend on the choice of an integer matrix $\bA$ specifying the coefficients of the linear combinations that should be decoded. 
However, finding the optimal choice of $\bA$ for GM-IF, AM-SIF and GM-SIF
appears to be a hard problem for which no efficient approximation algorithm is known, making these receivers currently infeasible to implement in practice. Nevertheless, results obtained in \cite{Bakoury.2015:ImpactofChannelVariation} by exhaustive search demonstrate that optimal GM-IF and AM-SIF significantly outperform AM-IF in terms of achievable rates, while GM-SIF outperforms all of them.

Our main contribution in this paper is to develop low-complexity optimization methods for these receivers (GM-IF, AM-SIF and GM-SIF) which can closely approach the performance obtained by an optimal search. Each proposed method is based on optimizing a more tractable lower bound on the achievable rate, which can be performed very efficiently.
Simulation results \blue{show} that the proposed method for GM-IF has a small gap compared to optimal performance, while the remaining two methods, at least for the scenarios tested, have performance almost indistinguishable from the optimal one. 

Another contribution of this paper is to develop a low-complexity implementation of these receivers using finite-length codes over a low-order constellation, in order to validate the information-theoretic results under practical constraints. Attaining the GM performance requires the use of full-diversity codes, from which we have adopted root \blue{low-density parity-check} (LDPC) codes. Simulation results are shown to be consistent with the theoretical ones, with an expected gap due to the finite codeword length.

The remainder of the paper is organized as follows. The system model is described in Section~\ref{sec:system-model}. Section~\ref{sec:integer-forcing} reviews integer forcing for static fading as well as block fading, while Section~\ref{sec:successive-integer-forcing} reviews successive integer forcing, again for both fading types. In Section~\ref{sec:proposed-methods}, we present our proposed methods for selecting $\bA$. Section~\ref{sec:practical-codes} describes our low-complexity implementation under practical constraints. Simulation results are shown in Section~\ref{sec:simulation-results}, covering the information-theoretic performance as well as the performance with practical codes. Lastly, our conclusions are presented in Section~\ref{sec:conclusion}.

\subsection{Notation}

For any $x>0$, define $\log^+(x) \triangleq \max(\log(x), 0)$. We denote row vectors as lowercase bold letters (e.g., $\bfx$) and matrices as uppercase bold letters (e.g., $\bfX$). The $\ell_2$-norm of a vector $\bx$ is denoted by $\|\bx\|$. The matrix $\bfX^\tr$ denotes the transpose of $\bfX$. We use $\bI$ and $\bzero$, respectively, to denote an identity and an all-zero matrix of appropriate size, which should always be clear from the context. The set of all $m \times n$ matrices with entries from the set $\calA$ is denoted $\calA^{m \times n}$.

\section{System Model} \label{sec:system-model}

\newcommand{\f}[1]{_{(#1)}}

Consider a discrete-time, real Gaussian MIMO \blue{multiple-access channel} (MAC) with $N_T$ single-antenna transmitters and one $N_R$-antenna receiver, subject to block fading with $F$ independent fading realizations per codeword.

Specifically, let $n$ be the codeword length and assume, for simplicity, that $F$ divides $n$. For $\ell=1,\ldots,N_T$, let $\bx_\ell \in \RR^n$ denote the vector transmitted by the $\ell$th transmitter, which can be represented by the $\ell$th row of a matrix $\bX \in \RR^{N_T \times n}$. Similarly, let $\bY \in \RR^{N_R \times n}$ be a matrix whose $j$th row represents the vector received by the $j$th receive antenna. Assuming fading realizations of equal length, the received matrix can be expressed as
\begin{equation}
\bY = \mat{\bY\f{1} & \cdots & \bY\f{F}}
\end{equation}
where, for $i=1,\ldots,F$,
\begin{equation}
\label{eq:channel-model}
\bY\f{i} = \bH\f{i}\bX\f{i} + \bZ\f{i}
\end{equation}
$\bH_{(i)} \in \RR^{N_R \times N_T}$ is the matrix of channel fading coefficients for the $i$th block, $\bX\f{i} \in \RR^{N_T \times (n/F)}$ is such that 
\begin{equation}
\bX = \mat{\bX\f{1} & \cdots & \bX\f{F}}
\end{equation}
and $\bZ\f{i} \in \RR^{N_R \times (n/F)}$ is a Gaussian noise matrix with i.i.d.\ entries of zero mean and variance $\sigma^2$. Note that $\bH\f{i}$ remains constant throughout the transmission of a block of $n/F$ symbols, but can vary independently between realizations.

For convenience, let
\begin{equation}
\bH\f{1:F} = \left( \bH\f{1},\ldots,\bH\f{F} \right).
\end{equation}
We assume that the receiver has perfect knowledge of the channel realization $\bH\f{1:F}$, while the transmitters do not have this knowledge and are only aware of the channel statistics.

Each transmitted vector $\bx_\ell$ is assumed to be the encoding of a message $\bw_\ell \in \calW$ produced by the $\ell$th transmitter, where $\calW$ is the message space. The encoder rate, which is the same for all transmitters, is defined as
\begin{equation}
R = \frac{1}{n}\log_2|\calW|.
\end{equation}
Moreover, the transmitted vectors must satisfy a (symmetric) power constraint
\begin{equation}
\frac{1}{n} \|\bx_\ell\|^2 \leq P.
\end{equation}
These assumptions on equal power and equal rate are reasonable since the transmitters do not have knowledge of the channel matrix. For convenience, we denote $\SNR = P/\sigma^2$.

At the receiver, the decoder attempts to recover all messages, producing estimates $\hat{\bw}_1,\ldots,\hat{\bw}_{N_T}$. An error is said to occur if $\hat{\bfw}_\ell \neq \bfw_\ell$ for any $\ell$. The error probability of the scheme (encoder/decoder pair) is $P_e = \EE[P_e(\bH\f{1:F})]$, where 
$P_e(\bH\f{1:F})$ denotes the error probability for a fixed channel realization.

For any fixed $\bH\f{1:F}$, a rate $R$ is said to be achievable if, for any $\epsilon,\delta > 0$ and sufficiently large~$n$, there exists a scheme of rate at least $R-\delta$ such that $P_e(\bH\f{1:F}) \leq \epsilon$.

For a given family of schemes (indexed by $n$), let $R_\textrm{scheme}(\bH\f{1:F})$ denote its maximum achievable rate under a fixed channel realization~$\bH\f{1:F}$. For a target rate $R$, the outage probability is defined as $p_\out(R) \triangleq \PP[R_\textrm{scheme}(\bH\f{1:F}) < R]$, and for a fixed probability $\rho \in (0,1]$, the outage rate is defined as $R_\out(\rho) \triangleq \sup \{R: p_\out(R) \leq \rho\}$.

\blue{
\medskip
\begin{remark}
We have adopted a real-valued channel in order to facilitate comparison with optimal IF methods that require exhaustive search---whose complexity becomes prohibitively large over a complex-valued channel even for low dimension---as well as with the existing literature, which mostly considers real-valued channels. There is no loss of generality, since it is always possible to express a complex-valued channel $\bfY_c = \bfH_c\bfX_c + \bfZ_c$ as a real-valued channel
\begin{equation*}
\mat{\Re(\bfY_c) \\ \Im(\bfY_c)} = \mat{\Re(\bfH_c) & -\Im(\bfH_c) \\ \Im(\bfH_c) & \Re(\bfH_c)}\mat{\Re(\bfX_c) \\ \Im(\bfX_c)} + \mat{\Re(\bfZ_c) \\ \Im(\bfZ_c)}.
\end{equation*}
Alternatively, all expressions presented here can be straightforwardly generalized to the complex case by replacing transpose with conjugate transpose.
\end{remark}
}

\section{Integer Forcing} \label{sec:integer-forcing}

To aid the understanding, we first review integer forcing for static fading ($F=1$), before describing its extension to block fading ($F \geq 2$). For simplicity, the superscript indicating the block index is omitted when $F=1$.

\subsection{Static Fading} \label{ssec:if-static}

An integer-forcing receiver \cite{Zhan.2014:IFLinearReceivers} shares the same basic structure of a conventional linear receiver, as illustrated in Fig.~\ref{fig:if}: 
\begin{figure}
\centering
\includegraphics[scale=1]{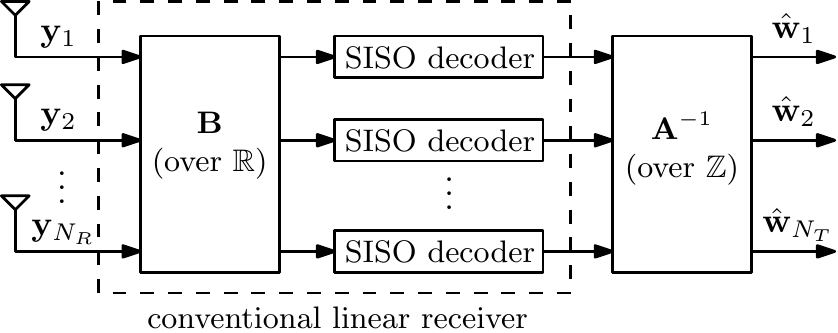}
\caption{Integer-forcing receiver.}
\label{fig:if}
\end{figure}
first, the received matrix is linearly transformed by an equalization matrix; then, each resulting stream is individually processed by a channel decoder. The difference, in the case of the integer-forcing receiver, is that the equalizer attempts to estimate not the transmitted signals directly but rather an integer linear transformation of the transmitted signals, which is then inverted after noise is removed.

Crucial to this noise removal step at the channel decoders is the use of a lattice code common to all transmitters \cite{Nazer.2011:CaFHarnessing,Zhan.2014:IFLinearReceivers}. A lattice $\Lambda \in \RR^n$ is a discrete subgroup of $\RR^n$, i.e., it is closed under integer linear combinations \cite{Zamir.2014:LatticeCoding}. In particular, a lattice can be expressed as $\Lambda = \{ \bx \in \RR^n : \bx = \bu \bG, \; \bu \in \ZZ^n \}$ where $\bG \in \RR^{n \times n}$ is the generator matrix of~$\Lambda$. It follows that if every $\bx_\ell \in \Lambda$, then $\ba \bX \in \Lambda$, for all $\ba \in \ZZ^{1 \times N_T}$ \cite{Zamir.2014:LatticeCoding}. In other words, since $\ba \bX$ is also a codeword from the same code $\Lambda$, then a decoder for $\Lambda$ is able to decode it (i.e., ``denoise'' it), regardless of the value chosen for $\ba$ \cite{Nazer.2011:CaFHarnessing}.

In more detail, consider the received matrix in \eqref{eq:channel-model},
\begin{equation}
\label{eq:channel-model-static}
\bY = \bH\bX + \bZ.
\end{equation}
Let $\bA \in \ZZ^{N_T \times N_T}$ be a full-rank integer matrix.
The receiver applies the equalization matrix $\bfB \in \RR^{N_T \times N_R}$ to create an effective channel output
\begin{align}
\bfY_\eff 
&= \bfB\bfY \\
&= \bfB\bfH\bfX + \bfB\bfZ \\
&= \bfA\bfX + \bfZ_\eff \\
&= \bV + \bfZ_\eff
\label{eq:effective-channel}
\end{align}
where
\begin{equation}
\bfZ_\eff = (\bfB\bfH - \bfA)\bfX + \bfB\bfZ
\label{eq:effective-noise}
\end{equation}
is the so-called \textit{effective noise} \cite{Zhan.2014:IFLinearReceivers} and
\begin{equation}
\bV = \bA \bX
\end{equation}
is an integer linear transformation of $\bX$.

Assuming that each $\bx_\ell$ is chosen from the same lattice $\Lambda \subseteq \RR^n$, we have that each row of
$\bV$
is also a lattice point from $\Lambda$ \cite{Nazer.2011:CaFHarnessing}. Thus, if decoding is successful and $\bV$ is recovered, then $\bX$ can also be recovered, provided $\bA$ is full-rank.\footnote{One might think that a stricter condition is required, namely, that $\bA$ is invertible. This would be true if any element of $\Lambda$ could be chosen for transmission without a power constraint. However, when nested lattice shaping is used to satisfy the power constraint, it is possible to show that requiring $\rank \bA = N_T$ is enough. See \cite{Zhan.2014:IFLinearReceivers,Nazer.2016:ExpandingComputeandForward} for details.}

As a consequence, it is shown in \cite{Nazer.2011:CaFHarnessing,Zhan.2014:IFLinearReceivers} that the following rate is achievable
\begin{equation}
\rif(\bH,\bA,\bB) \triangleq \min_m \frac{1}{2}\log^+\left(\frac{P}{\sigma_{\eff,m}^2}\right)
\end{equation}
where 
\begin{equation}
\sigma_{\eff,m}^2 = \norm{\bfb_m\bfH - \bfa_m}^2 P + \norm{\bfb_m}^2 \sigma^2
\label{eq:noise-variance}
\end{equation}
is the per-component variance of the vector $\bz_{\eff,m}$,
and $\ba_m$, $\bb_m$ and $\bz_{\eff,m}$ are the $m$th row of $\bA$, $\bB$ and $\bZ_{\eff}$, respectively.

The optimal equalization matrix $\bB$, for a given integer matrix $\bfA$, can be found using MMSE estimation \cite{Zhan.2014:IFLinearReceivers} as
\begin{equation}
\bfB = \bfA\bfH^\tr(\SNR^{-1}\bfI + \bfH\bfH^\tr)^\inv
\label{eq:equalization-matrix}
\end{equation}
In this case, we have \cite{Feng.2013:AlgebricApproach}
\begin{equation}
\sigma_{\eff,m}^2 = \sigma^2 \bfa_m\bfM\bfa_m^\tr
\label{eq:noise-modification}
\end{equation}
where
\begin{equation}
\bfM = (\SNR^{-1}\bfI + \bfH^\tr\bfH)^\inv
\label{eq:lattice-generator}
\end{equation}
and the achievable rate becomes
\begin{equation}\label{eq:rate-static}
\rif(\bH,\bA) \triangleq \min_m \frac{1}{2}\log^+\left(\frac{\SNR}{\bfa_m\bfM\bfa_m^\tr}\right).
\end{equation}

Optimizing the choice of $\bA$, we have the achievable rate
\begin{equation}\label{eq:rate-static-optimal}
\rif(\bH) \triangleq \max_{\bA: \rank \bA = N_T} \rif(\bH,\bA).
\end{equation}

As can be seen, the optimal matrix $\bfA$ solving \eqref{eq:rate-static-optimal} consists of a set of linearly independent integer vectors minimizing \eqref{eq:noise-modification}. Since $\bM$ is symmetric and positive definite \cite{Feng.2013:AlgebricApproach}, it admits a Cholesky decomposition $\bM = \bG \bG^\tr$, where $\bG \in \RR^{N_T \times N_T}$, leading to
\begin{equation}\label{eq:noise-variance-SVP}
\sigma_{\eff,m}^2 = \sigma^2 \| \bfa_m \bG \|^2.
\end{equation}
Thus, the optimal solution can be described as the integer coefficients (under the basis $\bG$) of a set of $N_T$ shortest linearly independent vectors in the lattice generated by $\bG$. This is known as the Shortest Independent Vector Problem (SIVP) \cite{Blomer.2000:ClosestVectors}, which is believed to be NP-Hard \cite{Regev.2005:OnLatticeRandomLinearCodesandCryptography}. However, suboptimal algorithms for basis reduction\footnote{For the special case of $N_T = 2$, an optimal basis can be found very efficiently using Lagrange's algorithm \cite{Agrell.2002:ClosestPointSearch}.}
exist that can find an approximation in polynomial time, such as the \blue{Lenstra-Lenstra-Lovasz} (LLL) algorithm \cite{Lenstra.1982:FactoringPolynomials}. 

Note that, if $\bfA = \bfI$ is chosen, then the scheme reduces to MMSE equalization \cite{Zhan.2014:IFLinearReceivers}. Thus, integer forcing generalizes linear equalization, providing potentially higher achievable rates.

\subsection{Block Fading} \label{ssec:if-block}

In the case of block fading, a complication arises, since now each block of the transmitted matrix experiences a different channel realization \cite{Knopp.2000:CodingforBlockFading}. While it is possible to independently equalize \blue{each block} of the received matrix, the resulting integer linear transformation $\bA$ must be the same for all blocks, in order for the rows $\bV$ to remain lattice codewords \cite{Bakoury.2015:ImpactofChannelVariation}.

More precisely, for each $i$th block, let $\bA\f{i} \in \ZZ^{N_T \times N_T}$ and $\bB\f{i} \in \RR^{N_T \times N_R}$ be its corresponding integer and equalization matrices, respectively, and compute the effective channel output for the block as
\begin{align}
\bfY_{\eff,(i)} 
&= \bfB_{(i)}\bfY_{(i)} \\
&= \bfB_{(i)}\bfH_{(i)}\bfX_{(i)} + \bfB_{(i)}\bfZ_{(i)} \\
&= \bfA\f{i}\bfX_{(i)} + \bfZ_{\eff,(i)}
\label{eq:effective-channel-mod} \\
&= \bV\f{i} + \bfZ_{\eff,(i)}
\end{align}
where $\bV\f{i} = \bA\f{i} \bX\f{i}$ and
\begin{equation}\label{eq:effective-noise-block-fading}
\bfZ_{\eff,(i)} = (\bfB\f{i}\bfH\f{i} - \bfA\f{i})\bfX\f{i} + \bfB\f{i}\bfZ\f{i}.
\end{equation}
It follows that
\begin{equation}
\bY_{\eff} = \bV + \bZ_{\eff}
\end{equation}
where $\bY_{\eff}$, $\bV$ and $\bZ_{\eff}$ denote the horizontal concatenation of $\bY_{\eff,(i)}$, $\bV\f{i}$ and $\bZ_{\eff,(i)}$, respectively, for $i=1,\ldots,F$. 

However, we can see that, in general, the rows of 
\begin{equation}
\bV = \mat{\bA\f{1}\bX\f{1} & \cdots & \bA\f{F}\bX\f{F}}
\end{equation}
are not guaranteed to be lattice points, since we cannot generally express $\bV$ as an integer linear transformation of $\bX$. 
\blue{For instance, the first half of a codeword concatenated to the second half of another codeword is not guaranteed to form a codeword in a general code.}\footnote{Unless the code can be expressed as the Cartesian product of shorter-length codes. But this effectively violates the assumption of block fading, corresponding to static fading with a shorter codeword length.}

Thus, for integer forcing to work over a block fading channel, we should require all $\bA\f{i}$ to be equal \cite{Bakoury.2015:ImpactofChannelVariation},
\begin{equation}
\bA\f{i} = \bA, \qquad i=1,\ldots,F
\end{equation}
so that
\begin{equation}
\bV = \bA \bX.
\end{equation}

For a given $\bfA$, the optimal equalization matrix for each $i$th block can be obtained as
\begin{equation}
\bfB_{(i)} = \bfA\bfH_{(i)}^\tr\left(\SNR^\inv\bfI + \bfH_{(i)}\bfH_{(i)}^\tr\right)^\inv
\label{eq:equalization-modification}
\end{equation}
and the per-component variance of the vector $\bz_{\eff,m,(i)}$, the $m$-th row of the $\bfZ_{\eff,(i)}$, is given as
\begin{equation}\label{eq:noise-variance-mod}
\sigma_{\eff,m,(i)}^2 = \sigma^2 \bfa_m\bfM_{(i)}\bfa_m^\tr
\end{equation}
where 
\begin{equation}
\bfM_{(i)} =  \left(\SNR^\inv\bfI + \bfH_{(i)}^\tr\bfH_{(i)}\right)^\inv
\label{eq:lattice-generator-mod}
\end{equation}
and
$\ba_m$ is the $m$th row of $\bA$.

Let $\bz_{\eff,m} = \mat{\bz_{\eff,m,(1)} & \cdots & \bz_{\eff,m,(F)}}$ denote the $m$th row of $\bZ_{\eff}$. Since the noise variance \eqref{eq:noise-variance-mod} may now be different for each block, an achievable rate expression is not immediate to obtain and may, in fact, depend on the type of decoder used \cite{Bakoury.2015:ImpactofChannelVariation}. Specifically, it depends on whether the decoder properly exploits the diversity inherent in the block fading channel.

Two decoding methods are proposed and analyzed in \cite{Bakoury.2015:ImpactofChannelVariation}: the Arithmetic Mean (AM) and the Geometric Mean (GM) decoders discussed below.

\subsubsection{AM decoder}

This decoder does not attempt to exploit the channel variation and instead treats each component of the effective noise vector $\bz_{\eff,m}$ as having the same variance \cite{Bakoury.2015:ImpactofChannelVariation}, denoted by $\sigma_{\textrm{AM},m}^{2}$. This variance is given by the \textit{arithmetic mean} (hence the decoder name) of the variance of the effective noise on each block,
\begin{equation}
\sigma_{\textrm{AM},m}^{2} = \frac{1}{F}\sum_{i=1}^{F} \sigma_{\eff,m,(i)}^{2}.
\label{eq:noise-AM}
\end{equation}

It is shown in \cite{Bakoury.2015:ImpactofChannelVariation} that the AM-IF receiver achieves the following rate
\begin{align}
\ramif(\bfH_{(1:F)},\bfA) &= \min_m \frac{1}{2}\log^+\left(\frac{P}{\sigma_{\textrm{AM},m}^{2}}\right) \nonumber \\
&= \min_m \frac{1}{2}\log^+\left(\frac{\SNR}{\bfa_m\bfM_{\textrm{AM}}\bfa_m^\tr}\right)
\label{eq:rate-AM}
\end{align}
where
\begin{equation}
\bfM_{\textrm{AM}} = \frac{1}{F} \sum_i \bfM_{(i)}.
\label{eq:lattice-generator-AM}
\end{equation}

Thus, the maximum rate achievable by this method is \begin{equation}\label{eq:rate-AM-optimal}
\ramif(\bH) \triangleq \max_{\bA: \rank \bA = N_T} \ramif(\bfH_{(1:F)},\bA).
\end{equation}

Note that \eqref{eq:rate-AM} is identical to \eqref{eq:rate-static} with $\bM$ replaced by $\bfM_{\textrm{AM}}$. Thus, \eqref{eq:rate-AM-optimal} can be solved in the same way as in the case of static fading  \cite{Bakoury.2015:ImpactofChannelVariation}. In particular, the LLL algorithm (or similar) may be used to find an approximately optimal solution in polynomial time.

A special case of AM-IF consists of choosing $\bA = \bI$, which corresponds to conventional MMSE equalization followed by an AM decoder \cite{Bakoury.2015:ImpactofChannelVariation}. This scheme is referred to as AM-MMSE and its achievable rate denoted as
\begin{equation}
R_{\textrm{AM-MMSE}}(\bfH_{(1:F)}) = \ramif(\bfH_{(1:F)}, \bfI).
\end{equation}

\subsubsection{GM decoder}

This decoder optimally exploits the fact that the effective noise variance is not constant across the blocks \cite{Bakoury.2015:ImpactofChannelVariation}. The rate achievable by this method can be understood as the average achievable rate among all the individual blocks, as if they could be treated as parallel channels (which is clearly an upper bound). More precisely, the rate achievable by GM-IF is proven in \cite{Bakoury.2015:ImpactofChannelVariation} to be
\begin{align}
\rgmif(\bfH_{(1:F)},\bfA) 
&= \min_m \frac{1}{F} \sum_i \frac{1}{2}\log^+\left(\frac{P}{\sigma_{\eff,m,(i)}^{2}}\right) \label{eq:rate-GM-addition}\\
&= \min_m \frac{1}{2}\log^+\left(\frac{P}{\sigma_{\mathrm{GM},m}^2}\right)
\label{eq:rate-GM}
\end{align}
where 
\begin{align}
\sigma_{\textrm{GM},m}^2 
&= \left(\prod_i \sigma_{\eff,m,(i)}^{2}\right)^{\frac{1}{F}} \label{eq:noise-GM} \\
&= \sigma^2 \cdot \left(\prod_i \bfa_m\bfM_{(i)}\bfa_m^\tr \right)^{\frac{1}{F}}.
\end{align}

Note that \eqref{eq:noise-GM} is the \textit{geometric mean} (hence the decoder name) of the variance of the effective noise in each block.

Therefore, the maximum achievable rate with GM-IF is
\begin{equation}\label{eq:rate-GM-optimal}
\rgmif(\bH) \triangleq \max_{\bA: \rank \bA = N_T} \rgmif(\bfH_{(1:F)},\bA).
\end{equation}

It is useful pointing out that, for the same matrix $\bfA$, the GM decoder always achieves a rate at least as high as that of the AM decoder, since $\sigma^2_{\mathrm{GM}} \leq \sigma^2_\mathrm{AM}$ due to AM-GM inequality \cite{Steele.2004:CauchySchwarzMasterClass}. However, there is currently no known efficient method to find an optimal (or approximately optimal) solution for $\bfA$ in \eqref{eq:rate-GM-optimal} \cite{Bakoury.2015:ImpactofChannelVariation}, making optimal GM-IF currently infeasible to implement in practice, especially as $N_T$ grows.

As before, a special case of GM-IF consists of choosing $\bA = \bI$, which corresponds to conventional MMSE equalization followed by a GM decoder \cite{Bakoury.2015:ImpactofChannelVariation}. This scheme is referred to as GM-MMSE and its achievable rate denoted as
\begin{equation}
R_{\textrm{GM-MMSE}}(\bfH_{(1:F)}) = \rgmif(\bfH_{(1:F)}, \bfI).
\end{equation}

\newcommand{\WhiteNoiseMatrix}{\bN}
\newcommand{\WhiteNoiseVector}{\bn}

\section{Successive Integer-Forcing} \label{sec:successive-integer-forcing}

One way to improve the performance of a conventional linear receiver is to apply successive interference cancellation (SIC) \cite{Wolniansky.1998:VBLAST}: after a codeword is successfully decoded, the receiver can use it as side information in order to cancel part of the interference, reducing the variance of the effective noise. This principle can be applied to integer-forcing as well \cite{Ordentlich.2013:SuccessiveIFsumRate}. 
Specifically, in a successive integer-forcing (SIF) receiver, each integer linear combination of codewords that is successfully decoded is used to cancel its contribution to the effective noise affecting the remaining linear combinations, potentially enabling a higher achievable rate.

Note that SIF decoding must be done sequentially, in contrast to conventional IF, which may be done in parallel. Thus, the decoding order is relevant. We assume that decoding follows the index of $\ba_m$, $m=1,\ldots,N_T$. Thus, in contrast to conventional IF, the ordering of the rows of $\bA$ may have an impact on the achievable rates for SIF.

We start by reviewing SIF for static fading, followed by its extension to block fading.

\subsection{Static Fading}\label{ssec:sif-static}

Recall the effective channel \eqref{eq:effective-channel} described in Section~\ref{sec:integer-forcing}. The SIF receiver exploits the fact that the rows $\bZ_\eff$ are correlated in general and starts by performing a whitening transformation.

Consider the generalized covariance matrix of $\bfZ_\eff$, defined as
\begin{equation}
\bfK_{\bfZ_\eff} \triangleq \frac{1}{n}\EV[\bfZ_\eff\bfZ_\eff^\tr].
\end{equation}
Assuming that the optimal equalization matrix $\bB$ is used, it is possible to show that \cite{Ordentlich.2013:SuccessiveIFsumRate,Zhan.2014:IFLinearReceivers}
\begin{equation}
\bfK_{\bfZ_\eff} = \sigma^2\bfA\bfM\bfA^\tr \label{eq:generalized-covariance-sif}
\end{equation}
where $\bfM$ is defined in \eqref{eq:lattice-generator}.

Since $\bfK_{\bfZ_\eff}$ is a symmetric positive definite matrix \cite{Ordentlich.2013:SuccessiveIFsumRate}, it admits a Cholesky decomposition $\bfK_{\bfZ_\eff} = \bfL\bfL^\tr$, where $\bfL$ is a lower triangular matrix with strictly positive diagonal entries. Define
\begin{equation}
\WhiteNoiseMatrix \triangleq  \bfL^\inv\bfZ_\eff.
\label{eq:noise-sif}
\end{equation}
Note that the generalized covariance matrix of $\WhiteNoiseMatrix$ is the identity matrix \cite{Ordentlich.2013:SuccessiveIFsumRate}. 

The effective channel output \eqref{eq:effective-channel} can be rewritten as
\begin{equation}
\bfY_{\eff} = \bV + \bfL\WhiteNoiseMatrix
\label{eq:effective-channel-sif}
\end{equation}
where $\bV = \bfA\bfX$. Since $\bfL$ is a lower triangular matrix,
we have that
\begin{equation}
	\bfy_{\eff,m} = \bv_{m} + \sum_{j=1}^{m} \ell_{m,j}\WhiteNoiseVector_j
\end{equation}
where $\ell_{m,j}$ denotes the $(m,j)$ entry of $\bfL$. Note that $\bfy_{\eff,m}$ is not affected by $\WhiteNoiseVector_{m'}$ for any $m' > m$.

The decoder acts in each row of $\bfY_{\eff}$ in a successive way. Suppose that $\bfv_1$ is successfully recovered. Then we can compute
\begin{equation}
\WhiteNoiseVector_{1} = \frac{1}{\ell_{1,1}}(\bfy_{\eff,1} - \bfv_1)
\end{equation}
and remove its influence on the second row of $\bfY_{\eff}$, 
\begin{equation}
\by_{2}' \triangleq \bfy_{\eff,2} - \ell_{2,1}\WhiteNoiseVector_{1} = \bv_{2} + \ell_{2,2}\WhiteNoiseVector_2
\end{equation}
so that $\bv_2$ can be decoded in the presence of less noise.

Generalizing, suppose that $\bfv_1,\dotsc,\bfv_{m-1}$ have been successfully recovered, providing the estimates $\WhiteNoiseVector_{1},\dotsc,\WhiteNoiseVector_{m-1}$. We can remove the influence of this noise to obtain
\begin{equation}
\by_{m}' \triangleq \bfy_{\eff,m} - \sum_{j=1}^{m-1} \ell_{m,j}\WhiteNoiseVector_j = \bv_{m} + \ell_{m,m}\WhiteNoiseVector_m
\end{equation}
so that $\bv_m$ can be decoded under noise of variance $\ell_{m,m}^2$. Then, the corresponding noise vector can be estimated as
\begin{equation}
\WhiteNoiseVector_m = \frac{1}{\ell_{m,m}}(\by_m' - \bv_{m}).
\end{equation}

Proceeding this way, it can be shown that the following rate is achievable \cite{Ordentlich.2013:SuccessiveIFsumRate}
\begin{equation}\label{eq:sif-rate}
	R_{\SIF}(\bfH,\bfA) = \min_m \frac{1}{2}\log^+\left(\frac{P}{\ell_{m,m}^2}\right).
\end{equation}
By choosing the optimal $\bfA$, the maximum achievable rate is
\begin{equation}
	R_\SIF(\bfH) = \max_{\bfA: \rank \bfA = N_T}R_\SIF(\bfH,\bfA).
	\label{eq:rsif}
\end{equation}

However, finding the optimal matrix for the SIF decoder is a different (and harder) problem than that for the IF decoder. In particular, each row permutation of $\bA$ may give a different achievable rate. As shown in \cite{Ordentlich.2013:SuccessiveIFsumRate}, it is possible to restrict the choice of $\bfA$ to the class of unimodular matrices and the optimal solution is obtained by finding a Korkin-Zolotarev (KZ) basis for the lattice generated by $\bG \in \RR^{N_T \times N_T}$, obtained from the Cholesky decomposition of $\bM = \bG \bG^\tr$.

Finding a KZ basis for a lattice involves finding a shortest lattice vector and is therefore an NP-hard problem \cite{Agrell.2002:ClosestPointSearch}. Suboptimal algorithms can be used, for example, applying the LLL algorithm $N_T$ successive times, where in each iteration the dimension of the underlying lattice decreases \cite{Agrell.2002:ClosestPointSearch}.

Note that it is possible to choose $\bfA = \pi(\bfI)$, where $\pi(\bfI)$ denotes a row permutation of the identity matrix. In this case, the method reduces to conventional SIC decoding, which is referred to as MMSE-SIC. In principle, all possible permutations could be tested, however this quickly becomes unattractive as the number of users increases.
Heuristic methods \cite{Wolniansky.1998:VBLAST} can be applied to find a good decoding order, for instance, decoding first the user with the highest SNR (i.e., lowest $\sigma_{\eff,m}^2$), at the expense of some performance degradation.

\subsection{Block Fading} \label{ssec:sif-block}

As in the case of parallel IF, for successive IF in the block fading scenario it is required that $\bfA$ remain the same for all blocks so that the rows of $\bV$ in \eqref{eq:effective-channel-sif} are still lattice points and lattice decoding and subsequent inversion is possible. All the other steps are the same as for static fading applied separately to each $i$th block, namely: equalization by $\bB_{(i)}$ given in \eqref{eq:equalization-modification}, Cholesky decomposition of
\begin{align}
\bfK_{\bfZ_{\eff,(i)}}  
&= \frac{1}{n/F}\EV[\bfZ_{\eff,(i)}\bfZ_{\eff,(i)}^\tr] \nonumber \\
&= \sigma^2\bfA\bfM_{(i)}\bfA^\tr \nonumber \\
&= \bfL_{(i)}\bfL_{(i)}^\tr
\end{align}
where $\bfM_{(i)}$ is defined in \eqref{eq:lattice-generator-mod},
and successive noise cancellation and estimation from
\begin{align}
\by_{m,(i)}' 
&\triangleq \bfy_{\eff,m,(i)} - \sum_{j=1}^{m-1} \ell_{m,j}\WhiteNoiseVector_{j,(i)} \\
&= \bv_{m,(i)} + \ell_{m,m,(i)}\WhiteNoiseVector_{m,(i)}.
\end{align}

Note that the effective channel after cancellation can be expressed more simply as
\begin{equation}
\by_{m}' = \bv_{m} + \bz_{m}'
\end{equation}
where $\by_{m}'$ and $\bz_{m}'$ are the horizontal concatenation of $\by_{m,(i)}'$ and $\bz_{m,(i)}' = \ell_{m,m,(i)}\WhiteNoiseVector_{m,(i)}$, respectively, for $i=1,\ldots,F$. In particular, each $i$th block of the reduced effective noise $\bz_{m}'$ has a possibly different variance $\ell_{m,m,(i)}^2$.

For the lattice decoding step, either of the two decoding methods discussed before, namely AM and GM decoding, may be used \cite{Bakoury.2015:ImpactofChannelVariation}. Their generalization to the case of block fading is straightforward.

\subsubsection{AM decoder}

This decoder treats $\bz_{m}'$ as white noise of variance
\begin{equation}
\sigma_{\textrm{AM-SIF},m}^{2} = \frac{1}{F}\sum_{i=1}^{F} \ell_{m,m,(i)}^{2}.
\label{eq:noise-AM-SIF}
\end{equation}
It follows that the achievable rate of AM-SIF is given by
\begin{equation}
\ramsif(\bfH_{(1:F)},\bfA) \triangleq \min_m \frac{1}{2} \log^+\left(\frac{P}{\sigma_{\textrm{AM-SIF},m}^2}\right)
\end{equation}
which, after optimizing over $\bA$, becomes
\begin{equation}
\ramsif(\bfH_{(1:F)}) \triangleq \max_{\bfA: \rank \bfA = N_T} \ramsif(\bfH_{(1:F)},\bfA).
\end{equation}
However, there is currently no known efficient method to find an optimal or approximately optimal choice of $\bfA$, which should be chosen to minimize \eqref{eq:noise-AM-SIF}.

Note that, if we choose $\bfA = \pi(\bfI)$, then the method reduces to conventional SIC with an AM decoder. This method is referred to as AM-SIC and its achievable rate is given by
\begin{equation}
\ramsic(\bfH_{(1:F)}) \triangleq \max_\pi \ramsif(\bfH_{(1:F)},\pi(\bfI)).
\end{equation}

\subsubsection{GM decoder}

As before, this decoder attempts to optimally exploit the variation of the noise statistics across blocks.

The achievable rate for GM-SIF can be computed as the average achievable rate all blocks, given by \cite{Bakoury.2015:ImpactofChannelVariation}
\begin{align}
\rgmsif(\bfH_{(1:F)},\bfA) &= \min_m \frac{1}{F} \sum_i \frac{1}{2} \log^+\left(\frac{\SNR}{\ell_{m,m,(i)}^2}\right) \\
&= \min_m \frac{1}{2} \log^+\left(\frac{\SNR}{\sigma_{\textrm{GM-SIF},m}^2}\right)
\end{align}
where 
\begin{equation}
\sigma_{\textrm{GM-SIF},m}^2 = \left(\prod_i \ell_{m,m,(i)}^2\right)^{\frac{1}{F}}.
\end{equation}
Note that, due the AM-GM inequality, for the same matrix $\bfA$, the GM decoder achieves a rate at least as high as that of the AM decoder.

The maximum achievable rate for GM-SIF is then given by
\begin{equation}
	\rgmsif(\bfH_{(1:F)}) = \max_{\bfA: \rank \bfA = N_T} \rgmsif(\bfH_{(1:F)},\bfA).
\end{equation}
However, 
there is currently no known efficient method to find an optimal or approximately optimal solution for $\bfA$, making optimal GM-SIF infeasible to implement in practice. Even an exhaustive search is more costly for GM-SIF than for GM-IF, since now all row permutations of the same $\bA$ must be considered.

As a special case, it is always possible to choose $\bfA = \pi(\bfI)$, which corresponds to conventional SIC with a GM decoder. This method is referred to as GM-SIC and its achievable rate is given by
\begin{equation}
	\rgmsic(\bfH_{(1:F)}) = \max_\pi \rgmsif(\bfH_{(1:F)},\pi(\bfI)).
\end{equation}

\section{Proposed Methods} \label{sec:proposed-methods}

Although AM-SIF, GM-IF and GM-SIF all have higher achievable rates than AM-IF, the fact that no efficient algorithm is known to find even an approximately optimal $\bfA$ can undermine the practical applicability of these methods.

In this section, we propose four suboptimal, low-complexity methods for choosing the integer matrix $\bA$. The first two are applicable to GM-IF, the third is applicable to AM-SIF, while the fourth applies to GM-SIF.

\subsection{Proposed Method 1 (GM-IF)} \label{ssec:prop1}

Let
\begin{equation}
\bfA_\textrm{AM-IF} \triangleq \argmax_{\bA: \rank \bA = N_T} \ramif(\bH_{(1:F)},\bA)
\end{equation}
be the optimal matrix $\bA$ for AM-IF.
We propose to use either this matrix or the identity matrix, depending on which one gives the highest rate under GM-IF decoding.

Let $\calA_1 = \{\bfI, \bfA_\textrm{AM-IF}\}$. 
The rate achievable by this method is given by
\begin{equation}
	R_{\prop1}(\bfH_{(1:F)}) \triangleq \max_{\bfA \in \calA_1} \rgmif(\bfH_{(1:F)},\bfA).
\end{equation}

Note that, \blue{if matrix $\bfA_\AM$ 
is chosen, then a rate at least as high as that of AM-IF is achieved, since for the same choice of $\bA$, the GM decoder always outperforms the AM decoder. On the other hand, if the identity matrix is chosen, then the proposed method becomes the same as GM-MMSE and, therefore, achieves the same rate.} Thus, the proposed method achieves rates as high as both GM-MMSE and AM-IF.

The complexity of this method is dominated by that of finding an optimal matrix for AM-IF, which can be approximated in polynomial time with the LLL algorithm. Therefore, the complexity is the same as that of AM-IF. 
\subsection{Proposed Method 2 (GM-IF)} \label{ssec:prop2}

In addition to the choices discussed above, we propose to test also the optimal matrix $\bA_{(i)}$ for each $i$th block that would be obtained with IF under static fading, namely
\begin{equation}\label{eq:prop2-A}
\bfA_{(i)} 
\triangleq \argmax_{\bA: \rank \bA = N_T} \rif(\bH_{(i)},\bA) \end{equation}
for $i=1,\ldots,F$.

Let $\calA_2 = \{\bfI, \bfA_\textrm{AM-IF}, \bfA_{(1)}, \dotsc, \bfA_{(F)}\}$. The rate achievable by this method is given by
\begin{equation}\label{eq:prop2-R}
	R_{\prop2}(\bfH_{(1:F)}) \triangleq \max_{\bfA \in \calA_2} \rgmif(\bfH_{(1:F)},\bfA).
\end{equation}

Note that Proposed Method 1 chooses a matrix which may be ``reasonably good'' for all blocks simultaneously but which is not necessarily optimal for any block. Proposed Method 2 expands this choice by including matrices which are optimal for at least one block, even if it they are worse for the others blocks. A reasoning behind this approach is that, in contrast to the AM decoder, the GM decoder is not limited by the performance of the worst block, since individual rates are added in \eqref{eq:rate-GM-addition}.

The complexity of Proposed Method 2 is higher than that of Proposed Method 1 since it is necessary run the LLL algorithm $F+1$ times. Since the LLL algorithm can be done in polynomial time, this proposed method still viable in practice, especially for small $F$.

\subsection{Proposed Method 3 (AM-SIF)} \label{ssec:prop3}

Recall that the AM-SIF receiver correctly takes into account the fact that the effective noise matrix has a different generalized covariance matrix $\bfK_{\bfZ_{\eff,(i)}}$ for each block, using each corresponding $\bL_{(i)}$ for noise cancellation; only the lattice decoding step treats the reduced effective noise $\bz_{m}'$ as having equal variance $\sigma_{\textrm{AM-SIF},m}^{2}$ across blocks.

An upper bound on this variance can be obtained by 
treating each block of the effective noise matrix~$\bZ_{\eff}$ as having the same generalized covariance matrix, given by
\begin{align}
\bfK_{\bfZ_{\eff}}  
&= \frac{1}{n}\EV\left[\bfZ_{\eff}\bfZ_{\eff}^\tr\right] \nonumber \\
&= \frac{1}{n}\sum_i \EV\left[\bfZ_{\eff,(i)}\bfZ_{\eff,(i)}^\tr\right] \nonumber \\
&= \frac{1}{F}\sum_i \bfK_{\bfZ_{\eff,(i)}} \\
&= \frac{1}{F}\sum_i \sigma^2\bfA\bfM_{(i)}\bfA^\tr \\
&= \sigma^2\bfA\bfM\bfA^\tr
\end{align}
where $\bM = \frac{1}{F} \sum_i \bfM_{(i)}$.
From this point on, we can proceed similarly to the case of static fading (Section~\ref{ssec:sif-static}), obtaining a reduced effective noise $\bz'_m$ with variance $\ell_{m,m}^2$, where $\bL$ is a lower triangular matrix with positive diagonal entries given by the Cholesky decomposition $\bfK_{\bfZ_{\eff}} = \bL \bL^\tr$. It follows that $\sigma_{\textrm{AM-SIF},m}^{2} \leq \ell_{m,m}^2$, since a suboptimal noise cancellation scheme is used.\footnote{This upper bound can also be derived directly by diagonalizing the positive definite matrix $\bL \bL^\tr = \frac{1}{F}\sum_i \bL_{(i)} \bL_{(i)}^\tr$.}

To be clear, the scheme described above uses block IF equalization, followed by static noise cancellation (SNC) and AM decoding, in contrast to AM-SIF, which uses block noise cancellation. To distinguish it from AM-SIF, we refer to this scheme as AM-SIF-SNC. Its achievable rate is then given by
\begin{equation}\label{eq:am-sif-snc-rate}
R_{\textrm{AM-SIF-SNC}}(\bfH_{(1:F)},\bfA) = \min_m \frac{1}{2}\log^+\left(\frac{P}{\ell_{m,m}^2}\right)
\end{equation}
which is a lower bound on $R_{\textrm{AM-SIF}}(\bfH_{(1:F)},\bfA)$.

Let
\begin{equation}
\bfA_\textrm{AM-SIF-SNC} \triangleq \argmax_{\bA: \rank \bA = N_T} R_{\textrm{AM-SIF-SNC}}(\bH_{(1:F)},\bA)
\end{equation}
be the optimal matrix $\bA$ for AM-SIF-SNC. We propose to use this matrix for AM-SIF. 
The rate achievable by this method is given by
\begin{equation}
	R_{\prop3}(\bfH_{(1:F)}) \triangleq \ramsif(\bfH_{(1:F)},\bA_\textrm{AM-SIF-SNC}).
\end{equation}

Note that optimizing \eqref{eq:am-sif-snc-rate} is exactly the same problem as optimizing \eqref{eq:sif-rate}. Thus, as discussed in Section~\ref{ssec:sif-static}, $\bfA_\textrm{AM-SIF-SNC}$ can be computed by KZ reduction of the lattice generated by $\bG \in \RR^{N_T \times N_T}$, obtained from the Cholesky decomposition of $\bM = \bG \bG^\tr$, and this procedure can be approximated by applying the LLL algorithm $N_T$ times.

\subsection{Proposed Method 4 (GM-SIF)} \label{ssec:prop4}

We can extended the same ideas of Proposed Method~2 to the case of successive decoding, simply by redefining \eqref{eq:prop2-A} and \eqref{eq:prop2-R} with their SIF counterparts, while using $\bA_\textrm{AM-SIF-SNC}$ instead of $\bA_\textrm{AM-SIF}$, as in Proposed Method 3. However, since the decoding order is important, in principle we would have to test all permutations of the identity matrix, but that number of permutations grows exponentially with the number of users. To avoid this complexity, we simply exclude the identity matrix (and all its permutations) from the set of possible choices for~$\bfA$.

Let $\calA_4 = \{\bfA_\textrm{AM-SIF-SNC}, \bfA_{\textrm{SIF},(1)}, \dotsc, \bfA_{\textrm{SIF},(F)}\}$, where,
for $i=1,\ldots,F$,
\begin{equation}
\bfA_{\textrm{SIF},(i)} \triangleq \argmax_{\bA: \rank \bA = N_T} \rsif(\bH_{(i)},\bA)
\end{equation}
is the optimal matrix for the $i$th block that would be obtained with SIF under static fading. 

The rate achievable by this method is given by
\begin{equation}
	R_{\prop4}(\bfH_{(1:F)}) = \max_{\bfA \in \calA_4} \rgmsif(\bfH_{(1:F)},\bfA).
\end{equation}

Similarly to Proposed Method 2, the complexity of this method grows linearly with the number of blocks. However, finding $\bfA_\textrm{AM-SIF-SNC}$ as well as each $\bfA_{\textrm{SIF},(i)}$ requires running the LLL algorithm $N_T$ times, for a total of $N_T (F+1)$ runs.

\subsection{Summary of Complexity}

\blue{Table~\ref{tab:complexity}
\begin{table}
\renewcommand{\arraystretch}{1.3}
\caption{Complexity of Integer-Forcing Methods}
\label{tab:complexity}
\centering
\begin{tabular}{cc}
\toprule
Method & Complexity \\
\midrule
AM/GM-MMSE & - \\
AM-IF & $\calO(N_T^4\log(N_T))$ \\
Proposed Method~$1$ & $\calO(N_T^4\log(N_T))$ \\
Proposed Method~$2$ & $\calO((F+1)N_T^4\log(N_T))$ \\
GM/AM-SIC & $\calO(N_T!)$ \\
Proposed Method~$3$ & $\calO(N_T^4\log(N_T))$ \\
Proposed Method~$4$ & $\calO((F+1)N_T^4\log(N_T))$ \\
Optimal GM/AM-SIF & $\calO((\sqrt{1+\SNR})^{N_T})$ \\
\bottomrule
\end{tabular}
\end{table}

shows the complexity of finding $\bA$ for each method discussed. For AM/GM-MMSE methods, since $\bA = \bI$ is known a priori, the complexity is negligible. The AM-IF and Proposed Method~$1$ have the same complexity, corresponding to a single run of the LLL algorithm \cite{Liu.2016:EfficientIntegerSearch}. Note that, for fixed $F$, Proposed Method~$2$ approaches the complexity of AM-IF and Proposed Method~$1$. For the successive scenario, the optimal choice of $\bA$ in AM/GM-SIC is found by testing all permutations of the identity matrix. Proposed Method~$3$ and Proposed Method~$4$ have similar complexity as Proposed Method~$1$ and $2$, respectively. Finally, the complexity of optimal GM/AM-SIF by exhaustive search follows from the bound in \cite{Zhan.2014:IFLinearReceivers} (see also \cite{Zhan.2014:IFLinearReceivers,Liu.2016:EfficientIntegerSearch}).

Note that, besides finding $\bA$, the receiver operation includes also the tasks of equalization and channel decoding, whose complexity scales with the blocklength $n$ and is identical for each method in the same category (parallel or successive). Thus, the task of finding $\bA$ tends to take a smaller fraction of the overall decoding time as $n$ grows.
}
\section{Implementation with Practical Codes} \label{sec:practical-codes}
 
All the achievable rate results discussed above assume the use of lattice codes of asymptotically high dimension over an asymptotically large constellation. However, in practice, a finite-length code and a finite-order modulation must be used. In this section, we discuss how practical lattice encoders and decoders for an IF receiver may be implemented with low complexity. We focus on the use of binary LDPC codes with $2$-PAM modulation.

\subsection{Encoding and Decoding} \label{ssec:encode-decode}

We start with conventional IF; the extension to successive~IF is straightforward. 
To simplify the description, with a slight abuse of notation, we consider the finite field of size 2, denoted $\ZZ_2$, as a subset of the integers, $\ZZ_2 = \{0,1\} \subseteq \ZZ$. Suppose the $\ell$th user encodes its message as a codeword $\bc_\ell \in \calC$ from a linear $(n,k)$ block code $\calC$ over $\ZZ_2$. The codeword $\bc_\ell$ is then modulated into a vector $\bx_\ell \in \{-\frac{1}{2}, \frac{1}{2}\}^n$ from a 2-PAM constellation, computed as
\begin{equation}
\bx_\ell = \bc_\ell + \bd_\ell \bmod 2
\end{equation}
where $\bd_\ell \in \{-\frac{1}{2}, \frac{1}{2}\}^n$ is a (discrete) \emph{dither} vector 
independent from $\bx_\ell$ and known to the receiver, and the $x \bmod 2$ operation is applied element-wise and assumed to return a real-valued number in the interval $(-1,1]$.

Note that the dither vector must be used in order to reduce the transmit power from the $\{0,1\}$ constellation to the $\{-\frac{1}{2},\frac{1}{2}\}$ constellation, and it could be interpreted more simply as a modulation map. In this case, a simple choice would be $\bd_\ell = (-\frac{1}{2},\ldots,-\frac{1}{2})$. However, using a random dither uniformly distributed over $\{-\frac{1}{2}, \frac{1}{2}\}^n$ is more convenient for our purposes since, as we shall see, it makes the error probability independent from the transmitted codeword. 
As a consequence of the use of dithers, the transmitted vector $\bx_\ell$ is not a lattice point anymore, in contrast to the description is sections~\ref{sec:integer-forcing} and~\ref{sec:successive-integer-forcing}. However, dithers can be easily removed at the receiver with a simple modification, re-enabling the results of the those sections. More precisely, let $\bfC$ and $\bfD$ be matrices whose $\ell$th row is $\bfc_\ell$ and $\bfd_\ell$, respectively. The receiver computes the effective channel output as
\begin{align}
\bY_{\eff} &= \mat{\bB_{(1)}\bY_{(1)} & \cdots & \bB_{(F)}\bY_{(F)}} - \bA\bD \bmod 2 \\
&= \bA\bfC + \bfZ_{\eff} \bmod 2 \\
&= \bV + \bfZ_{\eff} \bmod 2
\label{eq:mod2-channel}
\end{align}
where $\bZ_\eff$ is defined as in Section~\ref{sec:integer-forcing} and $\bV = \bA\bC \bmod 2$. Note that each row $\bv_m$ of $\bV$ is a codeword from $\calC$. Thus, we recover the same effective channel \eqref{eq:effective-channel}, except for the mod-2 operation.\footnote{Actually, a modulo-lattice channel is required for the results in \cite{Zhan.2014:IFLinearReceivers}, which is omitted in the IF overview given here and in \cite{Zhan.2014:IFLinearReceivers}. Thus, some modulo operation must be used at the receiver if the transmitted vectors satisfy a power constraint.}

For $j=1,\ldots,n$, let $y_{\eff,m}[j]$, $v_m[j]$, $z_{\eff,m}[j]$ denote the $j$th component of $\by_{\eff,m}$, $\bv_m$ and $\bz_{\eff,m}$, respectively. In order to use a belief propagation decoder, it is necessary to compute the log-likelihood ratio ($\LLR$) at the channel output, defined as
\begin{equation}
\LLR[j] = \log\left(\frac{\PP[y_{\eff,m}[j] \mid v_m[j] = 0]}{\PP[y_{\eff,m}[j] \mid v_m[j] = 1]}\right)
\end{equation}
for $j=1,\ldots,n$, where $\PP(\cdot | \cdot)$ denotes conditional probability.

To simplify the decoder, the (non-Gaussian) effective noise component $z_{\eff,m}[j]$ is treated as a Gaussian random variable with zero mean and variance $\sigma_{\eff,m}^2[j]$, where
\begin{equation}
\sigma_{\eff,m}^2[j] = \begin{cases}
\sigma_{\AM,m}^2, & \text{for AM decoding} \\
\sigma_{\eff,m,(I[j])}^2, & \text{for GM decoding}
\end{cases}
\end{equation}
and $I[j]$ denotes the index of the block to which the $j$th component belongs. However, since \eqref{eq:mod2-channel} is a mod-$2$ channel, there is an infinite number of noise realizations that lead to any given channel output, resulting in the expression
\begin{equation}
\PP[y_{\eff,m}[j] | v_m[j]] = \sum_{k\in 2\ZZ} \exp\left(-\frac{(y_{\eff,m}[j] -v_m[j] - k)^2}{2\sigma_{\eff,m}^2[j]}\right).
\end{equation}

In order to further simplify the LLR computation, we can approximate the above expression by keeping only the largest term, corresponding to the element $v_m[j] + 2\ZZ$ nearest to $y_{\eff,m}[j]$ in Euclidean distance. It follows that
\begin{equation}
\LLR[j] \approx \begin{cases}
\displaystyle \frac{1+2y_{\eff,m}[j]}{2\sigma_{\eff,m}^2[j]}, & -1 < y_{\eff,m}[j] \leq 0 \\
\displaystyle \frac{1-2y_{\eff,m}[j]}{2\sigma_{\eff,m}^2[j]}, & 0 \leq y_{\eff,m}[j] \leq 1
\end{cases}
\end{equation}
or more simply
\begin{equation}
\LLR[j] \approx \frac{1-2|y_{\eff,m}[j]|}{2\sigma_{\eff,m}[j]^2}
\end{equation}
since $y_{\eff,m}[j] \in (-1,1]$.

\begin{figure}
	\centering
	\includegraphics[width=0.5\textwidth]{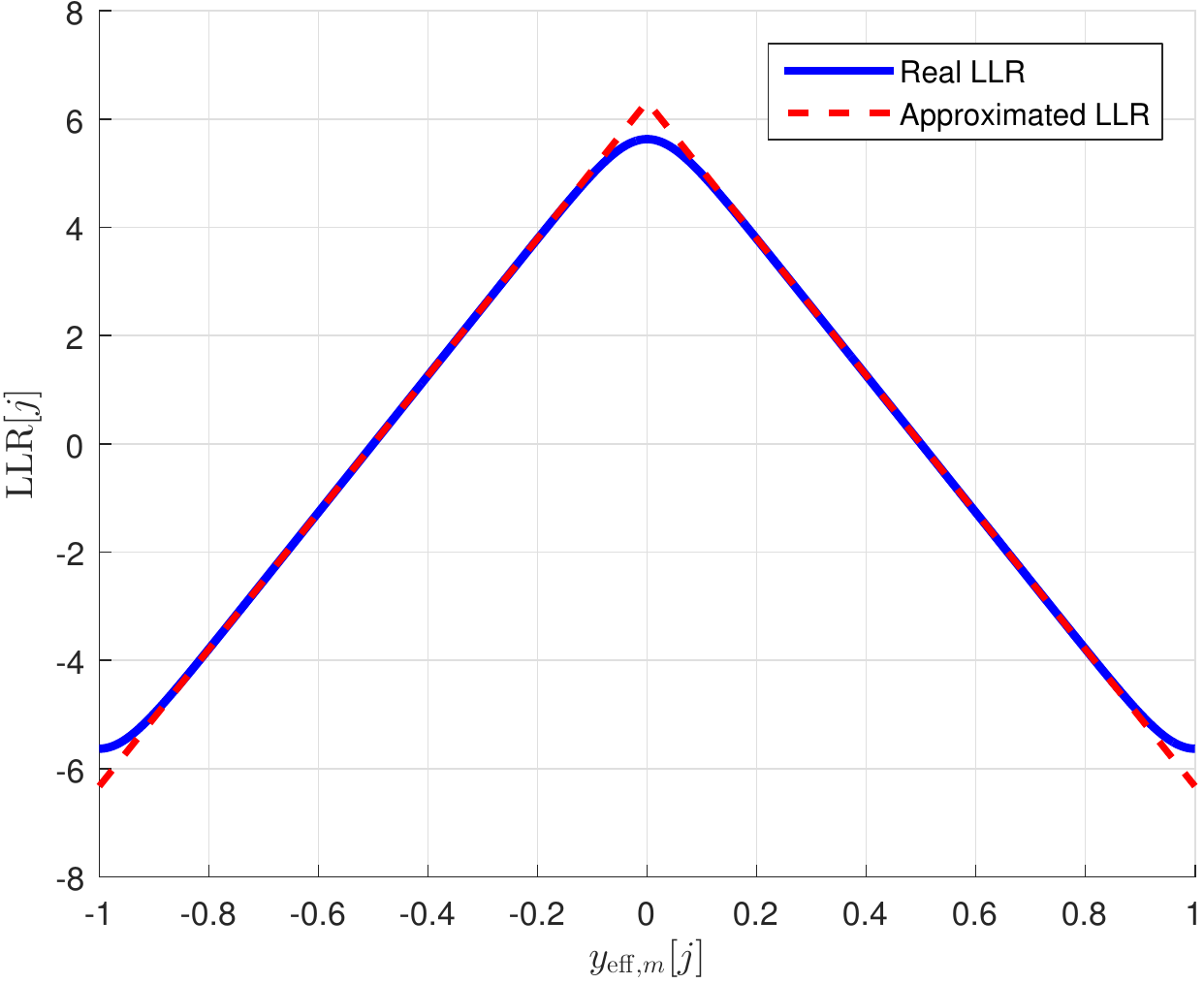}
	\caption{LLR in the mod-$2$ channel for $\SNR = 5$~dB.}
	\label{fig:LLR-mod2}
\end{figure}

Fig.~\ref{fig:LLR-mod2} 
shows the exact $\LLR$ in comparison with its approximation for a channel with $\SNR = 5$~dB. As we can see, for most of the input range the approximation is indistinguishable from the exact value. The approximation is slightly less accurate when 
the input
is close to an integer, but since these correspond to the peak LLR values, i.e., when there is a high degree of certainty about the value of $v_m[j]$, this loss of accuracy should not degrade the performance of belief propagation.

The above decoding procedure can be easily extended to successive IF by replacing $\by_{\eff,m}$ with $\by_m'$, $\bz_{\eff,m}$ with $\bz_m'$, and $\sigma_{\eff,m}^2[j]$ with
\begin{equation}
{\sigma}^2_{\SIF,m}[j] 
= \begin{cases}
\sigma_{\textrm{AM-SIF},m}^2, & \text{for AM decoding} \\
\ell_{m,m,(I[j])}^2, & \text{for GM decoding}.
\end{cases}
\end{equation}

\subsection{Code Construction} \label{ssec:code-construction}

In practice, approaching the GM performance (be it for MMSE, SIC, IF or SIF reception) requires not only a suitable decoder but also well-designed codes that allow the decoder to exploit diversity. This issue is not apparent in \cite{Bakoury.2015:ImpactofChannelVariation} since their achievable rate results (even for AM) are based on asymptotically good lattices that are already optimal for exploiting diversity. Designing such codes under finite-length and low-complexity constraints, however, is far from trivial.

It is well-known that an important parameter characterizing the performance of a code for a fading channel is its diversity order, defined as \cite{Biglieri.2005:CodingWirelessChannels}
\begin{equation}
d \triangleq - \lim_{\SNR \to \infty} \frac{\log P_e}{\log \SNR}
\end{equation}
where $P_e$ is the error probability of the decoder. For a block Rayleigh fading channel, the diversity order of a $q$-ary code is known to satisfy a Singleton-like bound \cite{Biglieri.2005:CodingWirelessChannels}
\begin{equation}
d \leq 1 + \left\lfloor F\left(1 - \frac{R}{\log_2 q}\right)\right\rfloor
\end{equation}
where 
$\lfloor \cdot \rfloor$ is the floor function and
$R$ is the code rate in bits per channel use.
Thus, codes that achieve full diversity ($d = F$) are limited by $R \leq (\log_2 q) / F$, or $R \leq 1/F$ for binary codes.

A family of rate-$1/F$ binary LDPC codes that achieve full diversity under belief propagation and have performance close to theoretical limits are the so-called \textit{root LDPC codes} \cite{Boutros.2010:LDPCforBlockFading}. 
These codes are systematic, with the information bits corresponding to the first $n/F^2$ positions of each block, and have a parity-check matrix $\bH \in \ZZ_2^{(F-1)n/F \times n}$ satisfying the following structure
\begin{equation}
\bH = \mat{\bH_{11} & \cdots & \bH_{1F} \\ \vdots & \ddots & \vdots \\ \bH_{F1} & \cdots & \bH_{F,F}}, \quad \bH_{ij} = \mat{\bH_{ij1} \\ \vdots \\ \bH_{i,j,F-1}}
\end{equation}
where
\begin{equation}
\bH_{ijk} = 
\begin{cases}
\bzero, & \text{$j<i$ and $k \neq j$} \\
\mat{\bI & \bzero}, & j=i \\
\bzero, & \text{$j>i$ and $k \neq j-1$} 
\end{cases}
\end{equation}
for all $1\leq i,j \leq F$ and all $1 \leq k \leq F-1$. This structure implies that, for each information bit from each $i$th block, there is one parity-check equation relating it to the bits from the $j$th block (and no other blocks), for all $j \neq i$.

It is worth mentioning that root LDPC codes under belief propagation guarantee full diversity only over the information bits.
Thus, if one is interested in recovering the entire codeword, it must be regenerated by re-encoding the information bits after decoding.

\section{Simulation Results} \label{sec:simulation-results}

In this section we present simulation results comparing the outage rate performance of our proposed methods with the optimal performance obtained by exhaustive search. For comparison, we include the performance of AM-IF (previously the only low-complexity IF receiver), as well as that of ML and conventional linear receivers. In our simulations, we specify an outage probability $\rho=0.01$, estimated by $10^4$ channel realizations. In each realization, the channel fading coefficients are drawn independently from a real-valued Gaussian distribution with zero mean and unit variance.

\begin{figure*}
	\centering
	\subfloat[]{\includegraphics[width=0.5\textwidth]{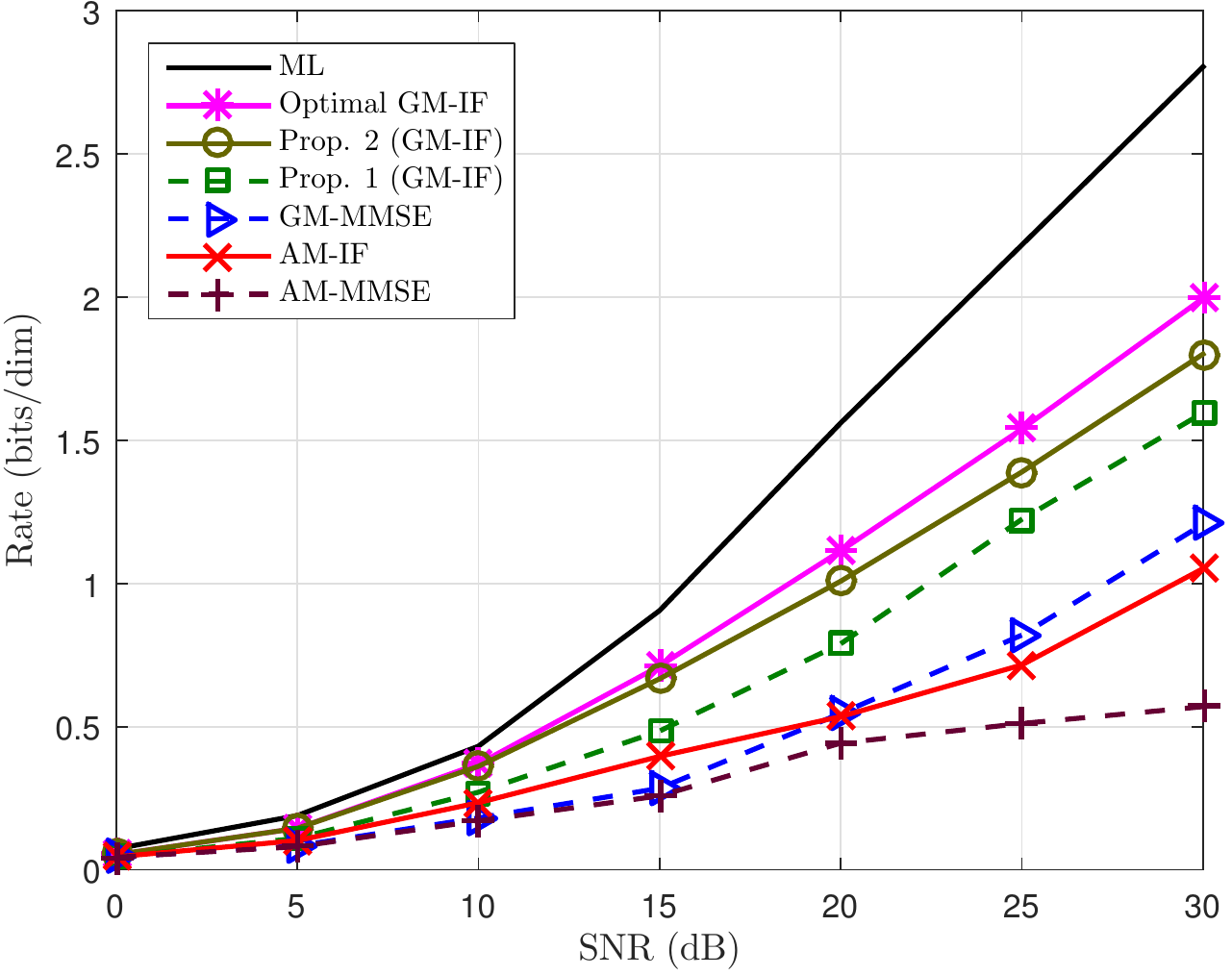}
		\label{fig:rate2x2x2-if}}	\subfloat[]{\includegraphics[width=0.5\textwidth]{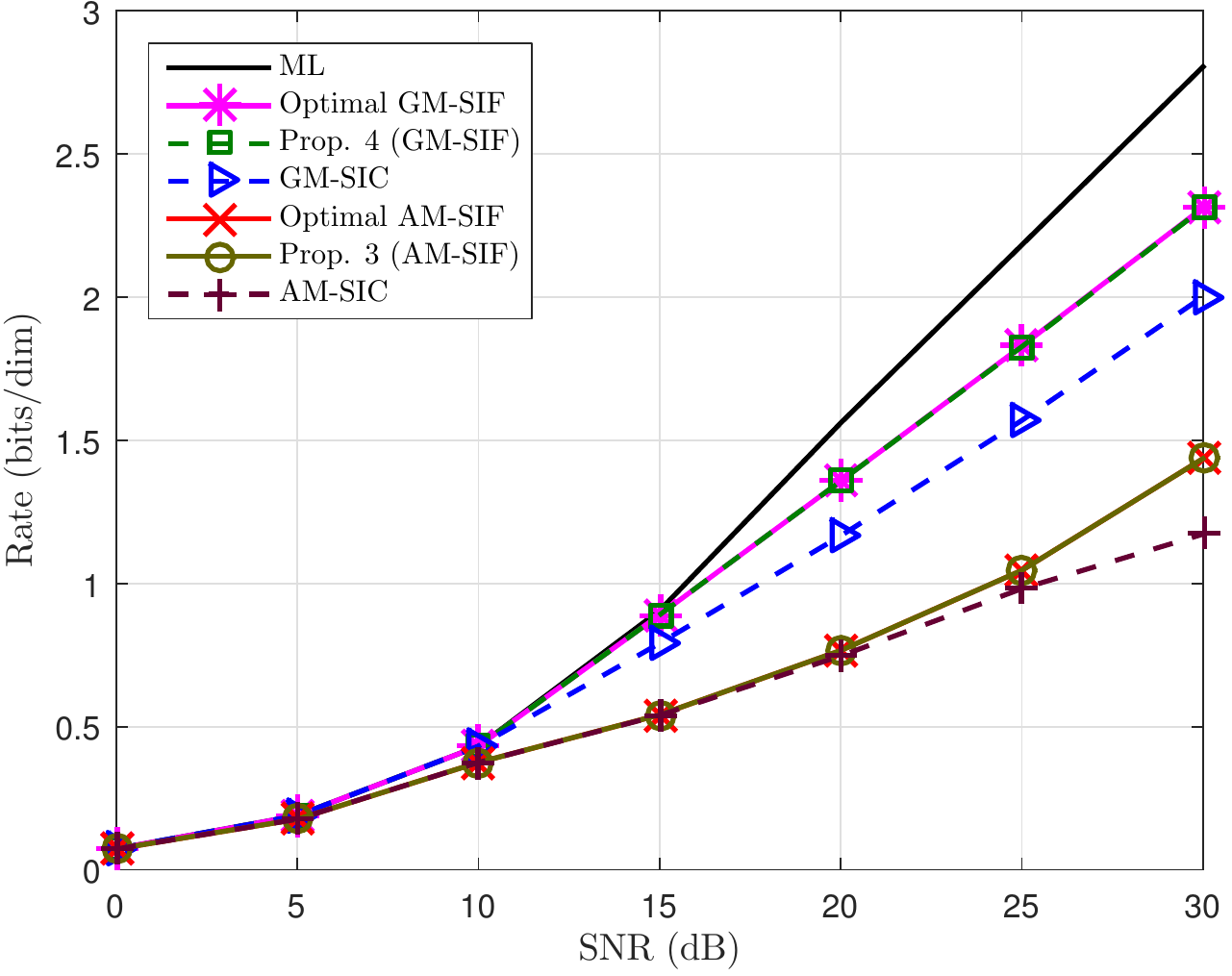}
		\label{fig:rate2x2x2-sif}}
	\caption{Outage rate on a $2\times2$ channel with $F=2$ blocks. \protect\subref{fig:rate2x2x2-if} Parallel decoding methods. \protect\subref{fig:rate2x2x2-sif} Successive decoding methods.}
	\label{fig:rate2x2x2}
\end{figure*}

For the optimal GM-IF, AM-SIF and GM-SIF, the matrix $\bA$ was obtained by an exhaustive search over all matrices vectors whose $\ell_1$-norm of each row does not exceed $15$. 
Fig.~\ref{fig:rate2x2x2} 
shows the outage rate for all these receivers on a $2~\times~2$ channel with $F=2$~blocks. As can be seen, Proposed Method 1 and 2 achieve performance close to optimal GM-IF and strictly higher than the maximum between AM-IF and GM-MMSE. In particular, for an outage rate of $1.5$~bits/dim, the performance of Proposed Method~$2$ is within $1.8$~dB of optimal GM-IF and outperforms Proposed Method~$1$ by $2.4$~dB. On the other hand, Proposed Method~$3$~and~$4$ appear to have performance indistinguishable from optimal AM-SIF and optimal GM-SIF, respectively. Note that Proposed Method~$4$ outperforms GM-SIC by approximately $3.2$~dB for an outage rate of $2$~bits/dim, 
while AM-SIF has a much lower performance in this case.

\begin{figure*}
	\centering
	\subfloat[]{\includegraphics[width=0.5\textwidth]{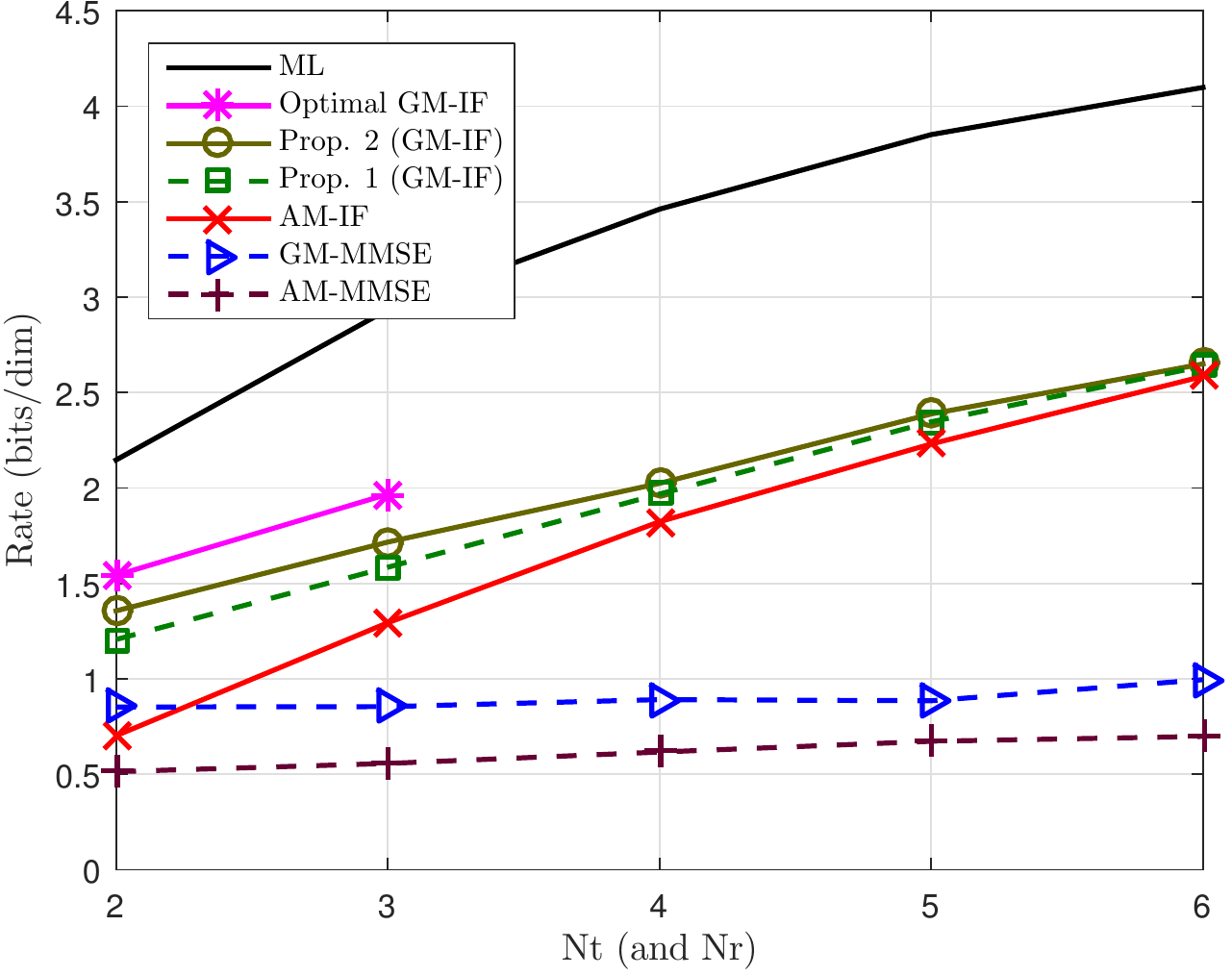}
		\label{fig:rateNtx2-if}}	\subfloat[]{\includegraphics[width=0.5\textwidth]{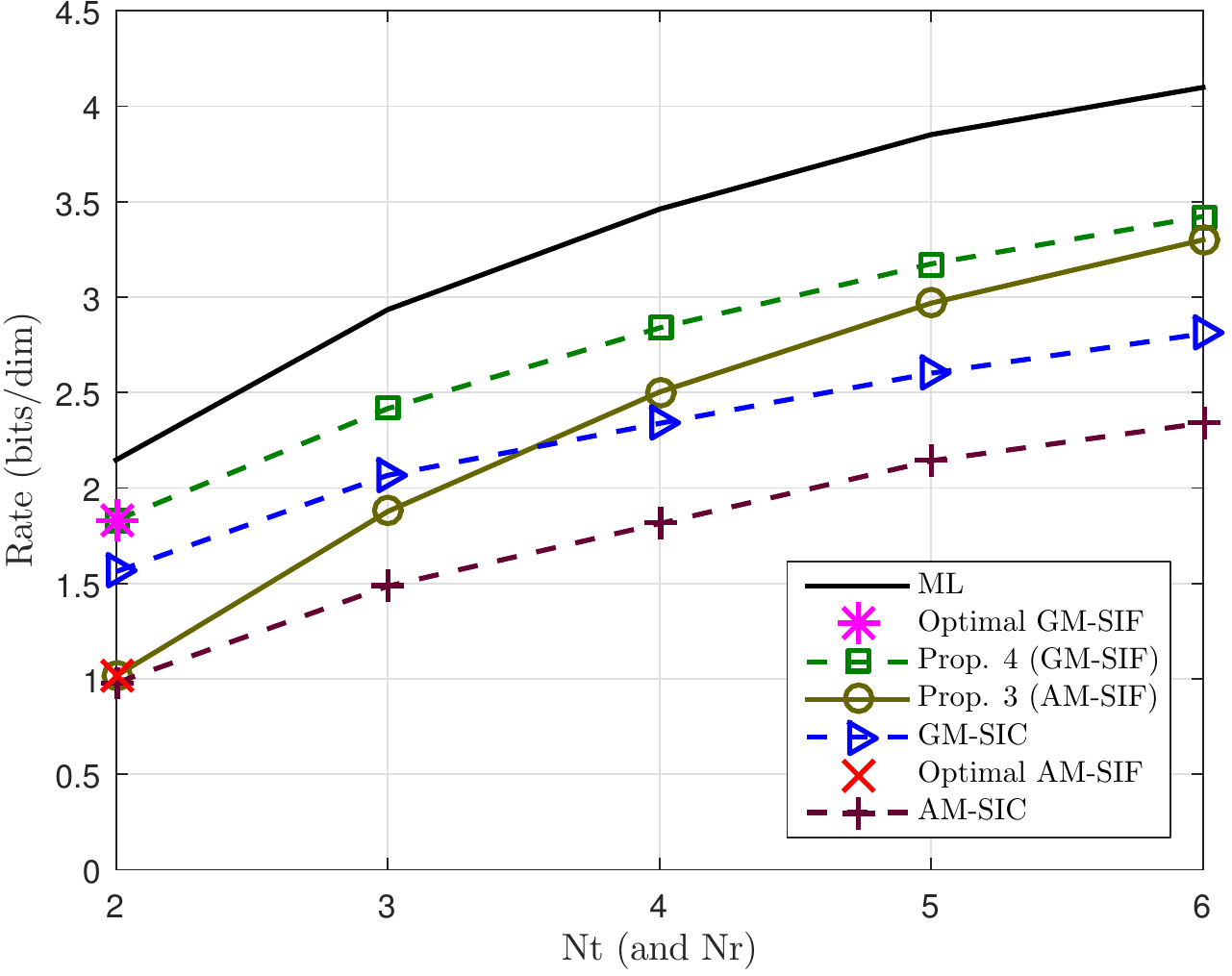}
		\label{fig:rateNtx2-sif}}
	\caption{Outage rate for a channel with $F=2$~blocks and $\SNR=25$~dB. \protect\subref{fig:rateNtx2-if} Parallel decoding methods. \protect\subref{fig:rateNtx2-sif} Successive decoding methods.}
	\label{fig:rateNtx2}
\end{figure*}

Fig.~\ref{fig:rateNtx2}
shows the outage rate for a scenario with $F=2$~blocks and $\SNR=25$~dB, varying the number of users $N_T$, while assuming the same number of receive antennas, $N_R = N_T$. Due to the complexity of exhaustive search, which grows exponentially with $N_T$, the performance of optimal GM-IF is shown only for $N_T=2$ and $N_T=3$, while that of optimal AM-SIF and GM-SIF is shown only for $N_T=2$. As can be seen, as $N_T$ increases, the performance of both Proposed Methods~$1$~and~$2$ appears to converge and is approached by that of AM-IF.
Similarly, the performance of Proposed Method~$3$ 
significantly improves as $N_T$ increases, 
outperforming GM-SIC for $N_T~\geq~4$ and approaching that of Proposed Method~$4$. 
Nevertheless, Proposed Method~$4$ still outperforms all other methods by a visible margin.

\begin{figure*}
	\centering
	\subfloat[]{\includegraphics[width=0.5\textwidth]{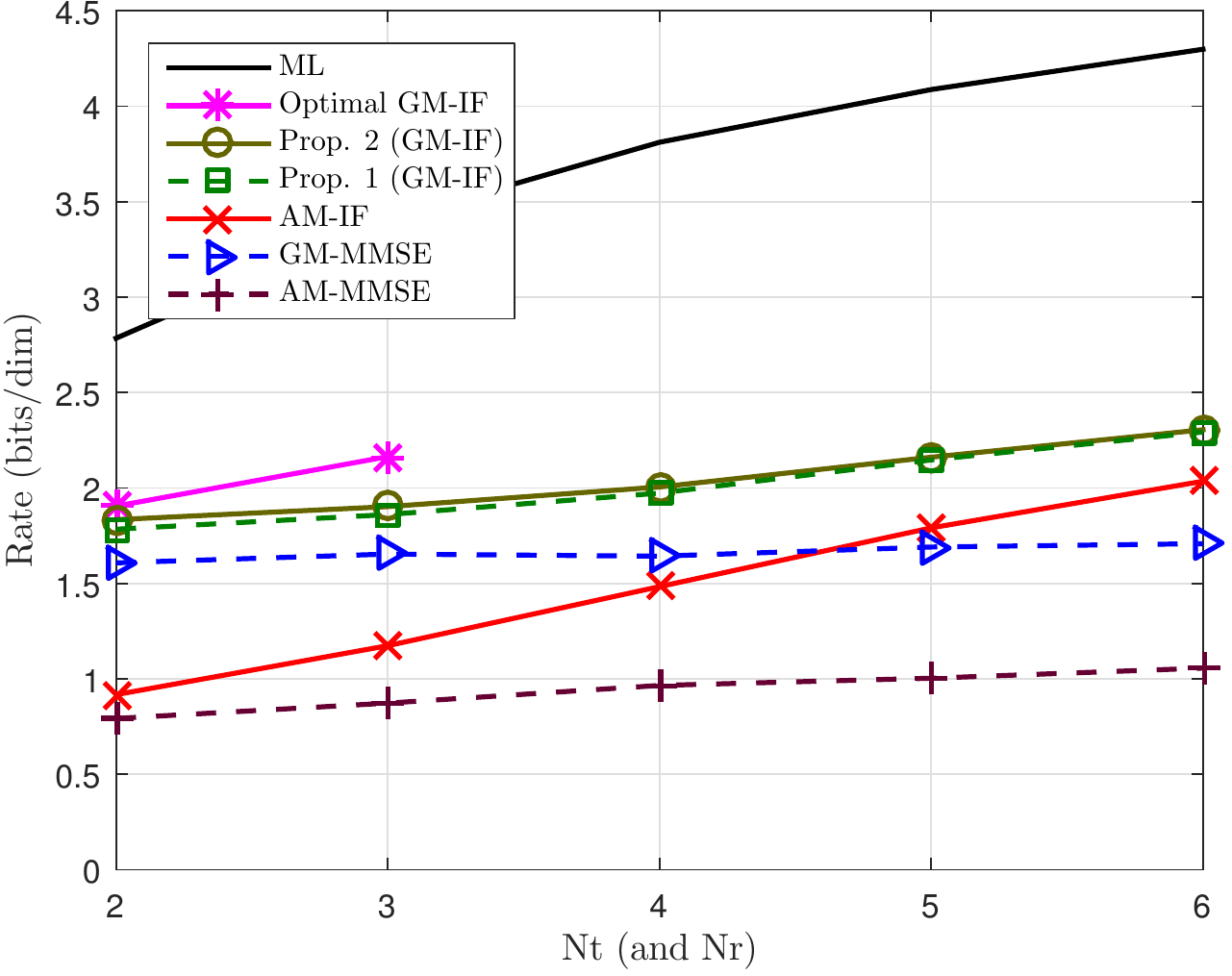} 
		\label{fig:rateNtx4-if}}	\subfloat[]{\includegraphics[width=0.5\textwidth]{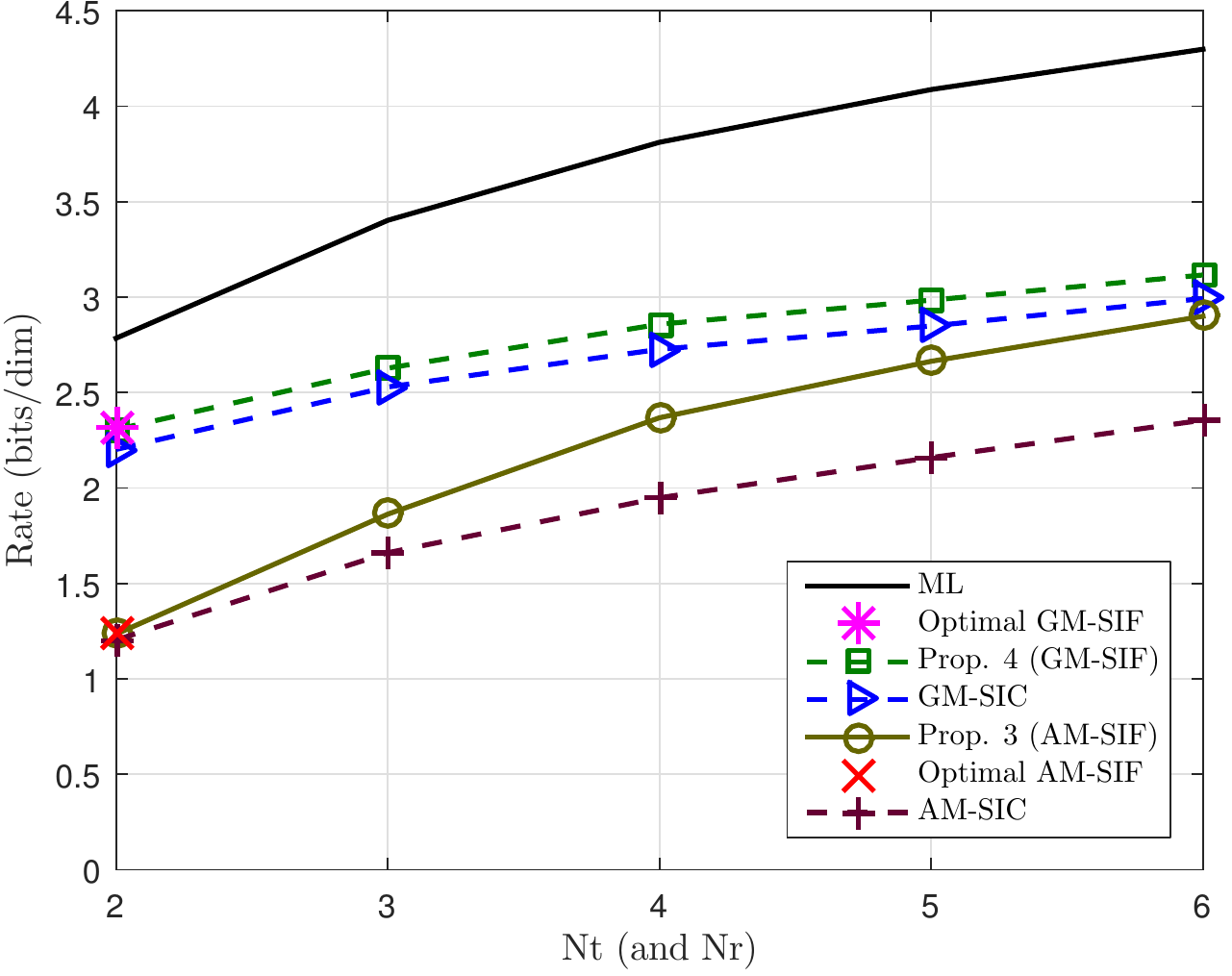} 
		\label{fig:rateNtx4-sif}}
	\caption{Outage rate for channel with $F=4$ blocks and $\SNR=25$~dB. \protect\subref{fig:rateNtx4-if} Parallel decoding methods. \protect\subref{fig:rateNtx4-sif} Successive decoding methods.}
	\label{fig:rateNtx4}
\end{figure*}

Fig.~\ref{fig:rateNtx4}
considers the same scenario as Fig.~\ref{fig:rateNtx2}, but with $F=~4$~blocks.
Similar observations can be made, 
except that, comparatively to Proposed Methods~1~and~2, the performance of AM-IF has worsened and that of GM-MMSE has improved, while still being significantly outperformed by the proposed methods.
\blue{
For the successive methods, a behavior similar to the $F=2$ case is observed, except that now GM-SIC outperforms Proposed Method~3 for all $N_T \leq 6$ and achieves a smaller gap to Proposed Method~4. While one might expect this gap to vanish for large $F$, it should be noted that, due to rate limitations, constructions of full-diversity codes are typically restricted to the small $F$ case.
}

\begin{figure}
	\centering
	\includegraphics[width=0.5\textwidth]{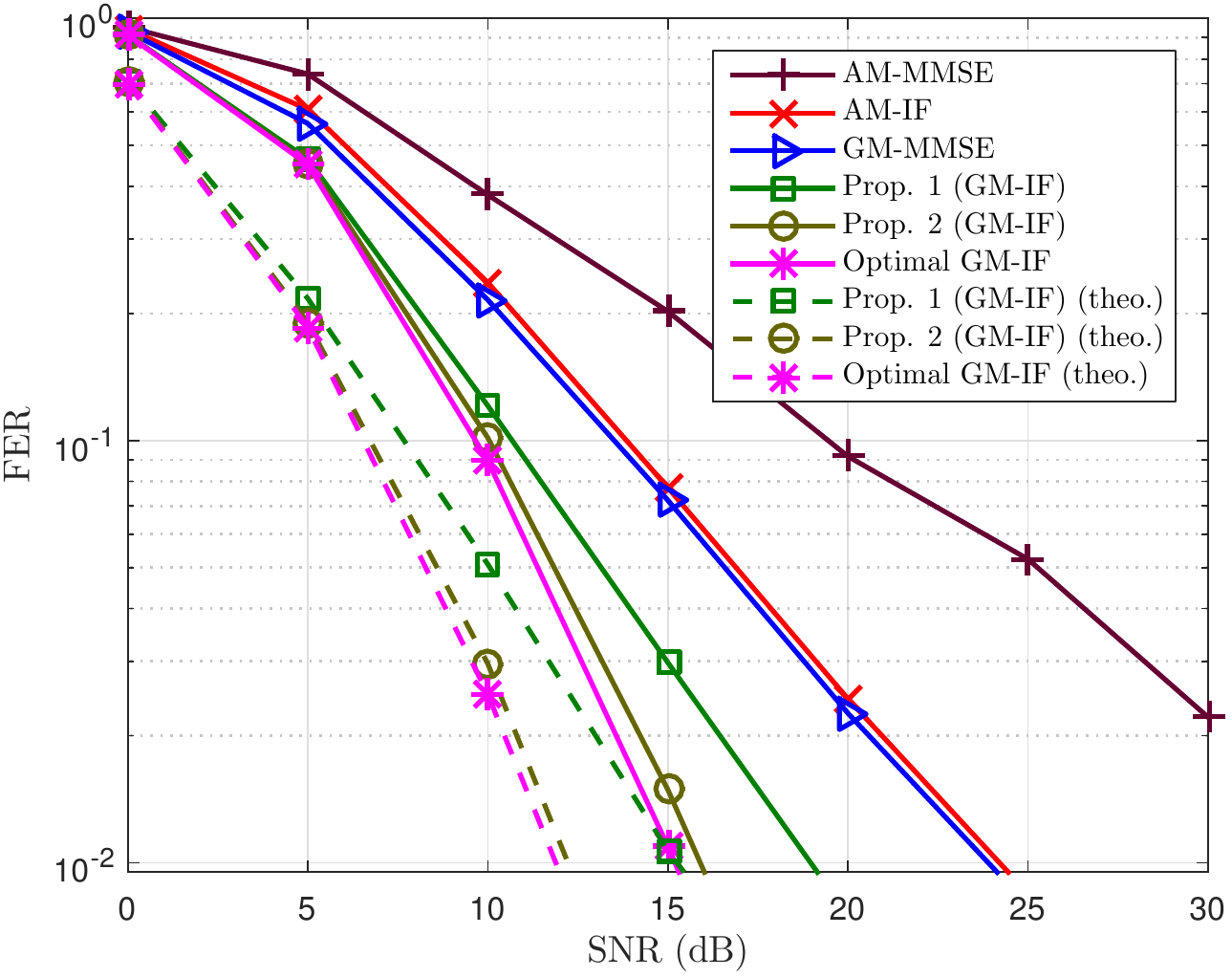}
	\caption{FER on a $2\times 2$ channel using parallel decoding methods. $R = 1/2$~bits/dim and $F=2$~blocks.}
	\label{fig:fer2x2x2-if}
\end{figure}

\begin{figure}
	\centering
	\includegraphics[width=0.5\textwidth]{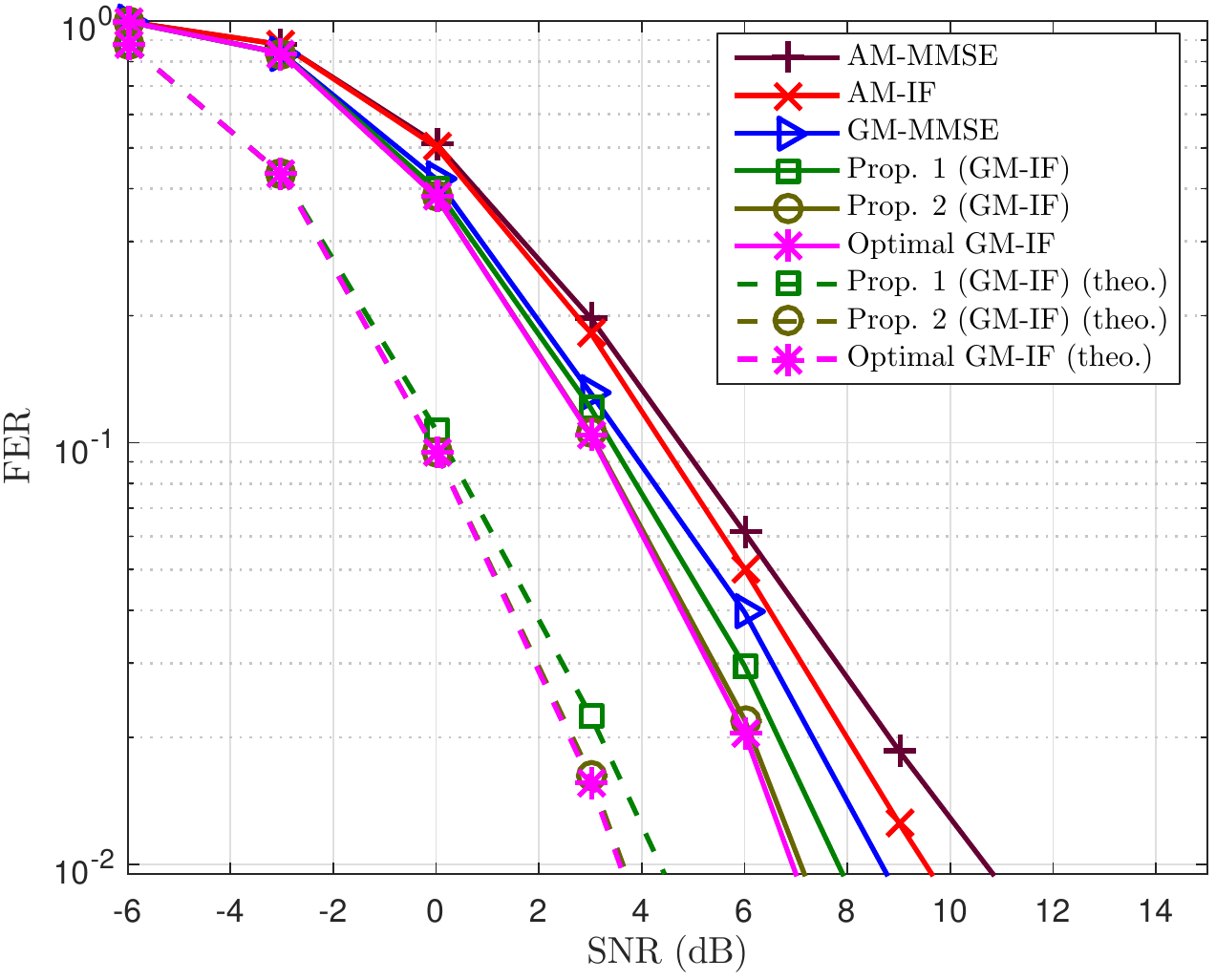}
	\caption{FER on a $2\times 2$ channel using parallel decoding methods. $R = 1/4$~bits/dim and $F=4$~blocks.}
	\label{fig:fer2x2x4-if}
\end{figure}

Lastly, Figs.~\ref{fig:fer2x2x2-if}~and~\ref{fig:fer2x2x4-if} show the frame-error rate (FER) on a $2~\times~2$ channel with $F=2$ and $F=4$, respectively, using 2-PAM modulation. A regular, rate-$1/F$ root-LDPC code \cite{Boutros.2010:LDPCforBlockFading} of length $n = 208$, constructed using a PEG-based technique \cite{Uchoa.2015:StructuredrootLDPCandPEGbasedforBlockFading}, is used in each simulation. As can be seen, the simulations are consistent with the theoretical results, with a performance gap due to the small constellation size and the non-optimality of the channel code. In particular, for a FER of $1\%$, both proposed methods are within $3.7$~dB of their theoretical FER for $F=2$ and within $3.4$~dB for $F=4$. 

The FER for successive decoding is not shown, since for $R \leq 1/2$ all methods have performance similar to their non-IF counterpart. For the benefits of SIF to become salient, lattice codes with higher spectral efficiency are needed.
The design of such codes, however, is outside the scope of this paper.
\section{Conclusions} \label{sec:conclusion}

In this paper, we propose four suboptimal methods for selecting an integer matrix $\bA$ for IF reception in a block fading scenario, two of them applicable to GM-IF and the other two applicable to AM-SIF and GM-SIF, respectively. 
The main idea behind these methods is to use a matrix $\bA$ optimized for a lower-performance scheme with a simpler objective function for which an approximately optimal solution can be found in polynomial time. 
For AM-SIF, the corresponding simpler scheme is the proposed AM-SIF-SNC scheme, while, for GM-IF and GM-SIF, it is the best choice among their respective AM counterpart and certain static fading solutions, all of which can be found very efficiently.

As shown by simulations, the proposed methods for GM-IF achieve outage rates strictly higher than both GM-MMSE and AM-IF (until now the best low-complexity methods), regardless of the number of blocks and users, while being only slightly more complex than \blue{AM-IF}. Exactly the same observations hold for GM-SIF in comparison with GM-SIC and AM-SIF-SNC.

We also show that AM-(S)IF and GM-(S)IF schemes can be realized in practice with low complexity, under finite codeword length and constellation constraints. Simulation results using full-diversity root LDPC codes 
are found to agree with theoretical ones, confirming the superiority of GM-IF in comparison with GM-MMSE and AM-IF.

An interesting avenue for future work is the development of low-complexity, full-diversity lattice codes with higher spectral efficiency 
(for instance, full-diversity $q$-ary linear codes with $q>2$ for use in Construction A \cite{Zhan.2014:IFLinearReceivers}).
Such codes would be directly applicable to the GM-IF and GM-SIF schemes, allowing a wider and more interesting operating range, in particular at outage rates for which these schemes are much superior to their AM or non-IF counterparts.

\section*{Acknowledgments}
The authors would like to thank Islam El Bakoury and Bobak Nazer for valuable discussions.

\begin{IEEEbiography}[{\includegraphics[width=1in,height=1.25in,clip,keepaspectratio]{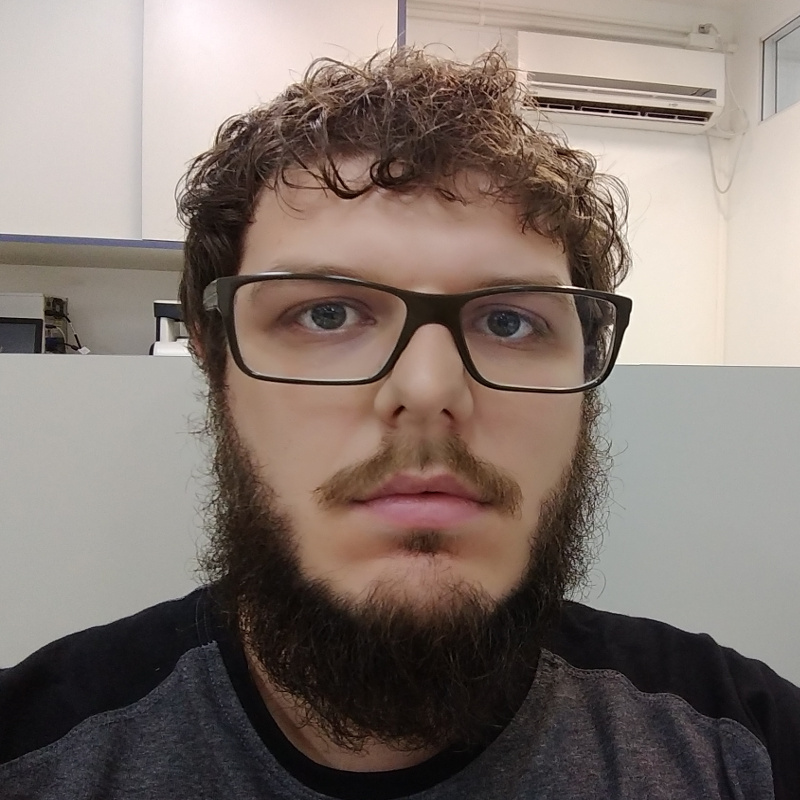}}]	{Ricardo Bohaczuk Venturelli} received the B.Sc.\ degree and M.Sc.\ degree from the Federal University of Santa Catarina (UFSC), Florian\'{o}polis, Brazil, in 2014 and 2016, respectively, all in electrical engineering. Currently, he is Ph.D. student in electrical engineering at the Federal University of Santa Catarina. His research interests include channel coding, network coding, information theory and wireless communications. He is a member of the Brazilian Telecommunications Society (SBrT) since~2014.
\end{IEEEbiography}
\newpage
\begin{IEEEbiography}[{\includegraphics[width=1in,height=1.25in,clip,keepaspectratio]{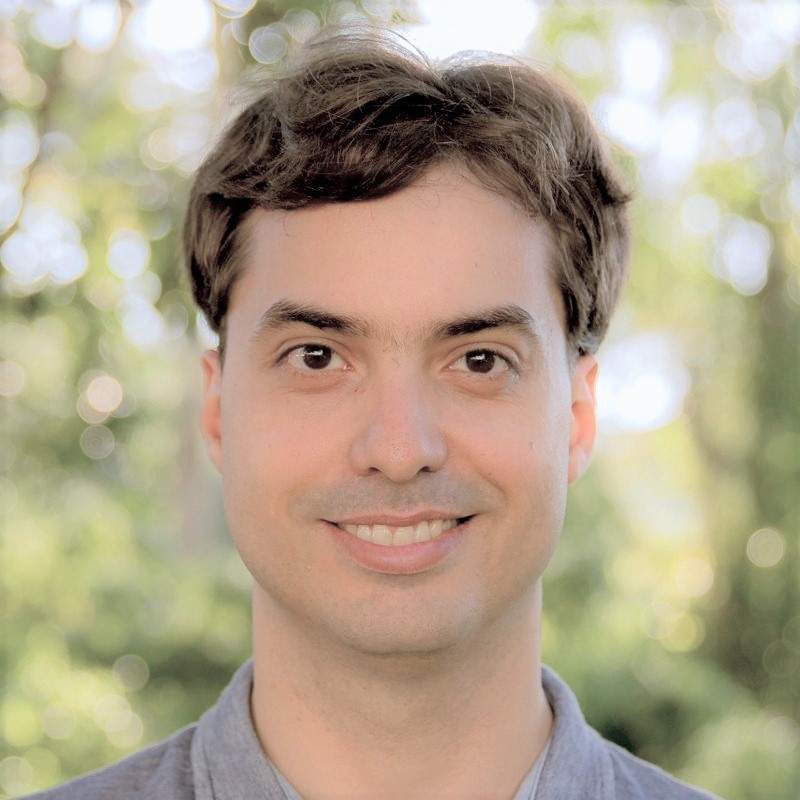}}]{Danilo Silva} (S'06--M'09) received the B.Sc.\ degree from the Federal University of Pernambuco (UFPE), Recife, Brazil, in 2002, the M.Sc.\ degree from the Pontifical Catholic University of Rio de Janeiro (PUC-Rio), Rio de Janeiro, Brazil, in 2005, and the Ph.D. degree from the University of Toronto, Toronto, Canada, in 2009, all in electrical engineering.
From 2009 to 2010, he was a Postdoctoral Fellow at the University of Toronto, at the \'Ecole Polytechnique F\'ed\'erale de Lausanne (EPFL), and at the State University of Campinas (UNICAMP).
In 2010, he joined the Department of Electrical Engineering, Federal University of Santa Catarina (UFSC), Brazil, where he is currently an Assistant Professor. His research interests include wireless communications, channel coding, information theory, and machine learning.

Dr. Silva is a member of the Brazilian Telecommunications Society (SBrT). He was a recipient of a CAPES Ph.D. Scholarship in 2005, the Shahid U. H. Qureshi Memorial Scholarship in 2009, and a FAPESP Postdoctoral Scholarship in 2010.
\end{IEEEbiography}

\end{document}

%% file: preamble.tex
\usepackage{amsmath,amssymb}

\newcommand{\rif}{R_{\textrm{IF}}}
\newcommand{\rsif}{R_{\textrm{SIF}}}
\newcommand{\ramif}{R_{\textrm{AM-IF}}}
\newcommand{\rgmif}{R_{\textrm{GM-IF}}}
\newcommand{\ramsif}{R_{\textrm{AM-SIF}}}
\newcommand{\rgmsif}{R_{\textrm{GM-SIF}}}
\newcommand{\ramsic}{R_{\textrm{AM-SIC}}}
\newcommand{\rgmsic}{R_{\textrm{GM-SIC}}}

\DeclareMathOperator{\rank}{\textsf{rank}}

\newcommand{\PP}{\mathbb{P}}
\newcommand{\EE}{\mathbb{E}}

\newcommand{\tr}{\mathrm{T}}
\newcommand{\eff}{\textrm{eff}}
\newcommand{\SNR}{\mathrm{SNR}}

\newcommand{\SIF}{\textrm{SIF}}

\newcommand{\LLR}{\textrm{LLR}}

\newcommand{\out}{\textrm{out}}
\newcommand{\inv}{{-1}}
\newcommand{\AM}{\textrm{AM}}

\newcommand{\prop}{\textrm{prop}}

\DeclareMathOperator*{\argmax}{\arg\max}

\newcommand{\EV}{\mathrm{E}}

\newcommand{\norm}[1]{\begin{Vmatrix} #1 \end{Vmatrix}}

\newenvironment{remark}{\textit{Remark: }}{}
\def\qed{\endIEEEproof}

\newcommand{\mat}[1]{\begin{bmatrix} #1 \end{bmatrix}}

\newcommand{\ZZ}{\mathbb{Z}}

\newcommand{\RR}{\mathbb{R}}

\renewcommand{\Re}{\mathsf{Re}}
\renewcommand{\Im}{\mathsf{Im}}

\newcommand{\bzero}{\boldsymbol{0}}

\newcommand{\calA}{\mathcal{A}}

\newcommand{\calC}{\mathcal{C}}

\newcommand{\calO}{\mathcal{O}}

\newcommand{\calW}{\mathcal{W}}

\newcommand{\bfA}{\mathbf{A}}
\newcommand{\bfB}{\mathbf{B}}
\newcommand{\bfC}{\mathbf{C}}
\newcommand{\bfD}{\mathbf{D}}

\newcommand{\bfH}{\mathbf{H}}
\newcommand{\bfI}{\mathbf{I}}

\newcommand{\bfK}{\mathbf{K}}
\newcommand{\bfL}{\mathbf{L}}
\newcommand{\bfM}{\mathbf{M}}

\newcommand{\bfX}{\mathbf{X}}
\newcommand{\bfY}{\mathbf{Y}}
\newcommand{\bfZ}{\mathbf{Z}}

\newcommand{\bfa}{\mathbf{a}}
\newcommand{\bfb}{\mathbf{b}}
\newcommand{\bfc}{\mathbf{c}}
\newcommand{\bfd}{\mathbf{d}}

\newcommand{\bfv}{\mathbf{v}}
\newcommand{\bfw}{\mathbf{w}}
\newcommand{\bfx}{\mathbf{x}}
\newcommand{\bfy}{\mathbf{y}}

\newcommand{\bA}{\mathbf{A}}
\newcommand{\bB}{\mathbf{B}}
\newcommand{\bC}{\mathbf{C}}
\newcommand{\bD}{\mathbf{D}}

\newcommand{\bG}{\mathbf{G}}
\newcommand{\bH}{\mathbf{H}}
\newcommand{\bI}{\mathbf{I}}

\newcommand{\bL}{\mathbf{L}}
\newcommand{\bM}{\mathbf{M}}
\newcommand{\bN}{\mathbf{N}}

\newcommand{\bV}{\mathbf{V}}

\newcommand{\bX}{\mathbf{X}}
\newcommand{\bY}{\mathbf{Y}}
\newcommand{\bZ}{\mathbf{Z}}

\newcommand{\ba}{\mathbf{a}}
\newcommand{\bb}{\mathbf{b}}
\newcommand{\bc}{\mathbf{c}}
\newcommand{\bd}{\mathbf{d}}

\newcommand{\bn}{\mathbf{n}}

\newcommand{\bu}{\mathbf{u}}
\newcommand{\bv}{\mathbf{v}}
\newcommand{\bw}{\mathbf{w}}
\newcommand{\bx}{\mathbf{x}}
\newcommand{\by}{\mathbf{y}}
\newcommand{\bz}{\mathbf{z}}